\documentclass[acmlarge]{acmart}
\AtBeginDocument{%
  }

\setcopyright{cc}
\setcctype{by-nc-nd}
\acmJournal{IMWUT}
\acmYear{2026} \acmVolume{10} \acmNumber{3} \acmArticle{130}
\acmMonth{9} \acmDOI{10.1145/3831978}

\usepackage{tikz}
\usepackage{amsmath}

\usepackage{tikz}
\usepackage{tabularx}
\usepackage{caption}
\usepackage{multirow}
\usepackage{multicol}
\usepackage{makecell}
\usepackage{adjustbox}
\usepackage{subfigure}
\usepackage{booktabs}

\usepackage{amsmath,amssymb,amsfonts}
\usepackage{filecontents}
\usepackage{listings}
\usepackage{multirow}
\usepackage{multicol}
\usepackage[inline]{enumitem}
\usepackage{xcolor}

\usepackage{acronym}
\usepackage{xspace,url}
\usepackage{bbding}
\usepackage{pifont}
\usepackage{wasysym}
\usepackage{tcolorbox}
\tcbuselibrary{breakable} %
\usepackage{colortbl}
\usepackage{enumitem} %
\setlist[itemize]{leftmargin=*,nosep}
\setlist[enumerate]{leftmargin=*,nosep}
\usepackage{diagbox}
\usepackage{hyperref}
\makeatletter
\def\UrlAlphabet{%
      \do\a\do\b\do\c\do\d\do\e\do\f\do\g\do\h\do\i\do\j%
      \do\k\do\l\do\m\do\n\do\o\do\p\do\q\do\r\do\s\do\t%
      \do\u\do\v\do\w\do\x\do\y\do\z\do\A\do\B\do\C\do\D%
      \do\E\do\F\do\G\do\H\do\I\do\J\do\K\do\L\do\M\do\N%
      \do\O\do\P\do\Q\do\R\do\S\do\T\do\U\do\V\do\W\do\X%
      \do\Y\do\Z}
\def\UrlDigits{\do\1\do\2\do\3\do\4\do\5\do\6\do\7\do\8\do\9\do\0}
\g@addto@macro{\UrlBreaks}{\UrlOrds}
\g@addto@macro{\UrlBreaks}{\UrlAlphabet}
\g@addto@macro{\UrlBreaks}{\UrlDigits}
\makeatother

\usepackage{pdflscape}
\usepackage{amsmath}

\usepackage{mathtools} %

\usepackage{stfloats}

\usepackage{cleveref}

\newcommand{\ie}{\textit{i.e.}\xspace}
\newcommand{\eg}{\textit{e.g.}\xspace}

\newcommand{\revise}[1]{{#1}}

\newcommand{\tool}{\textsc{GUIAuditor}\xspace}
\newcommand{\compreduce}{\textsc{Mod\ding{172}}\xspace}
\newcommand{\companalysis}{\textsc{Mod\ding{173}}\xspace}
\newcommand{\compstorage}{\textsc{Mod\ding{174}}\xspace}
\newcommand{\toolnaive}{\textsc{GUIAuditor}\textsubscript{naive}\xspace}
\newcommand{\tooluniform}{\textsc{GUIAuditor}\textsubscript{MineContext}\xspace}
\newcommand{\toolpf}{\textsc{GUIAuditor}\textsubscript{w/o (Pixel+Feature)}\xspace}
\newcommand{\toolf}{\textsc{GUIAuditor}\textsubscript{w/o Feature}\xspace}

\renewcommand{\S}{Section}

\usepackage{etoolbox}

\AtBeginEnvironment{thebibliography}{%
  \interlinepenalty=10000
  \clubpenalty=10000
  \widowpenalty=10000
}

\makeatletter
\let\GUIAuditor@titlecase@secfont\@secfont
\g@addto@macro\@secfont{\MakeUppercase}
\newenvironment{titlecasesections}
  {\let\@secfont\GUIAuditor@titlecase@secfont}
  {}
\makeatother

\title{\tool: Enabling Post-hoc Child Safety Forensics via Action-Guided GUI Provenance on Mobile Devices}

\begin{document}

\author{Junlin Liu}
\orcid{0000-0001-5917-2251}
\affiliation{
  \institution{Peking University}
  \country{China}
}
\email{jlinliu@pku.edu.cn}

\author{Yifeng Cai}
\orcid{0000-0002-0049-6670}
\affiliation{
  \institution{Peking University}
  \country{China}
}
\affiliation{
  \institution{Beijing Tongming Lake Information Technology Application Innovation Center}
  \country{China}
}
\email{caiyifeng@pku.edu.cn}

\author{Shuai Wang}
\orcid{0009-0002-6297-9085}
\affiliation{
  \institution{Peking University}
  \country{China}
}
\email{wangshuai\_2016@pku.edu.cn}

\author{Zhineng Zhong}
\orcid{0009-0007-2490-5671}
\email{zhongzhineng@pku.edu.cn}

\author{Shaofei Li}
\orcid{0009-0001-6530-5935}
\email{lishaofei@pku.edu.cn}

\author{Jiacheng Liu}
\orcid{0009-0004-3759-6247}
\affiliation{
  \institution{Peking University}
  \country{China}
}
\email{jiachengliu25@stu.pku.edu.cn}

\author{Yuanchun Li}
\orcid{0000-0002-1591-2526}
\affiliation{
  \institution{Tsinghua University}
  \country{China}
}
\email{liyuanchun@air.tsinghua.edu.cn}

\author{Ziqi Zhang}
\orcid{0000-0001-8493-0261}
\affiliation{
  \institution{University of Illinois Urbana-Champaign}
  \country{USA}
}
\email{ziqi24@illinois.edu}

\author{Xiangqun Chen}
\orcid{0000-0002-7366-5906}
\email{cherry@pku.edu.cn}

\author{Ding Li}
\orcid{0000-0001-7558-9137}
\email{ding\_li@pku.edu.cn}

\author{Yao Guo}
\orcid{0000-0001-5064-5286}
\affiliation{
  \institution{Peking University}
  \country{China}
}
\email{yaoguo@pku.edu.cn}
\authornote{Corresponding author.}

\authorsaddresses{%
Authors' Contact Information:
\href{https://orcid.org/0000-0001-5917-2251}{Junlin Liu}, MOE Key Lab of HCST (PKU), School of Computer Science, Peking University, China, \href{mailto:jlinliu@pku.edu.cn}{\textcolor{black}{jlinliu@pku.edu.cn}};
\href{https://orcid.org/0000-0002-0049-6670}{Yifeng Cai}, MOE Key Lab of HCST (PKU), School of Computer Science, Peking University, China, and Beijing Tongming Lake Information Technology Application Innovation Center, China, \href{mailto:caiyifeng@pku.edu.cn}{\textcolor{black}{caiyifeng@pku.edu.cn}};
\href{https://orcid.org/0009-0002-6297-9085}{Shuai Wang}, Peking University, China, \href{mailto:wangshuai_2016@pku.edu.cn}{\textcolor{black}{wangshuai\_2016@pku.edu.cn}};
\href{https://orcid.org/0009-0007-2490-5671}{Zhineng Zhong}, MOE Key Lab of HCST (PKU), School of Computer Science, Peking University, China, \href{mailto:zhongzhineng@pku.edu.cn}{\textcolor{black}{zhongzhineng@pku.edu.cn}};
\href{https://orcid.org/0009-0001-6530-5935}{Shaofei Li}, MOE Key Lab of HCST (PKU), School of Computer Science, Peking University, China, \href{mailto:lishaofei@pku.edu.cn}{\textcolor{black}{lishaofei@pku.edu.cn}};
\href{https://orcid.org/0009-0004-3759-6247}{Jiacheng Liu}, MOE Key Lab of HCST (PKU), School of Computer Science, Peking University, China, \href{mailto:jiachengliu25@stu.pku.edu.cn}{\textcolor{black}{jiachengliu25@stu.pku.edu.cn}};
\href{https://orcid.org/0000-0002-1591-2526}{Yuanchun Li}, Institute for AI Industry Research (AIR), Tsinghua University, China, \href{mailto:liyuanchun@air.tsinghua.edu.cn}{\textcolor{black}{liyuanchun@air.tsinghua.edu.cn}};
\href{https://orcid.org/0000-0001-8493-0261}{Ziqi Zhang}, Department of Computer Science, University of Illinois Urbana-Champaign, USA, \href{mailto:ziqi24@illinois.edu}{\textcolor{black}{ziqi24@illinois.edu}};
\href{https://orcid.org/0000-0002-7366-5906}{Xiangqun Chen}, MOE Key Lab of HCST (PKU), School of Computer Science, Peking University, China, \href{mailto:cherry@pku.edu.cn}{\textcolor{black}{cherry@pku.edu.cn}};
\href{https://orcid.org/0000-0001-7558-9137}{Ding Li}, MOE Key Lab of HCST (PKU), School of Computer Science, Peking University, China, \href{mailto:ding_li@pku.edu.cn}{\textcolor{black}{ding\_li@pku.edu.cn}};
\href{https://orcid.org/0000-0001-5064-5286}{Yao Guo} (corresponding author), MOE Key Lab of HCST (PKU), School of Computer Science, Peking University, China, \href{mailto:yaoguo@pku.edu.cn}{\textcolor{black}{yaoguo@pku.edu.cn}}%
}

\begin{abstract}

  The proliferation of smart devices exposes children to online risks like grooming and financial scams that are deeply embedded within legitimate applications. Current approaches rely on automated prevention and detection, a paradigm that is fundamentally limited by its inherent fallibility. Whether rule-based or AI-driven, they inevitably produce false positives and negatives, failing to provide reliable protection. In this paper, we argue for a complementary, human-in-the-loop, post-hoc forensic paradigm. We present \tool, the first system designed to realize this vision by creating GUI Provenance: a queryable, semantic record of a child's interaction sequence. To generate this, \tool leverages a Multimodal Large Language Model (MLLM) to translate the temporal sequence of GUI events into a human-understandable narrative. To make this practical on mobile devices, a novel evidence distillation pipeline reduces the data requiring analysis by over 89.2\% compared to periodic sampling approaches adopted by industry standards, with negligible impact on accuracy. On a new dataset of 295 interaction clips, \tool achieves a 95.23\% Macro-F1 Score in logging significant events and, crucially, its two-stage forensic query engine successfully retrieves the correct evidence as the top result for over 90.20\% of natural language questions. \revise{An end-to-end evaluation on three modern smartphones shows that the full pipeline, including on-device MLLM inference, adds 2.1W of power draw and 7.4s of per-event latency, with a peak memory footprint of ${\sim}$3.1GB.} These results show that post-hoc GUI forensics can run on modern mobile devices and provide useful context for guardian-led safety review.
  \end{abstract}



\renewcommand{\shortauthors}{Liu et al.}

\maketitle

\section{Introduction}
\label{sec:intro}
The proliferation of interactive smart devices, such as smartphones and tablets, has fundamentally reshaped the landscape of modern childhood. These devices serve as the primary portals through which children access the digital world, spanning education, entertainment, and social communication, all mediated by rich \acp{gui}. This pervasive digital immersion presents a double-edged sword. While it provides unprecedented access to information, creative tools, and social platforms, it simultaneously exposes children to a new frontier of complex risks deeply embedded within application ecosystems~\cite{cac-regulations,cybersafekids,european2023influence,di2020ui,livingstone20214cs}. For instance, as highlighted in a recent BBC report~\cite{bbc-gromming}, a child may be groomed on a children's gaming website by an adult predator who convincingly mimics teenage slang to build trust, ultimately causing severe real-world harm.

Current technical approaches to this problem, whether rule-based or AI-driven, have converged on the goal of automating risk prevention or detection. Direct rule-based controls, such as OS-level app blockers~\cite{apple-screen-time,google-family-link} and network filters~\cite{adguard}, are limited to coarse-grained boundary decisions and are therefore blind to the interactions within permitted applications. Broader system-level controls, like OS permission managers~\cite{alsoubai2022permission,tahaei2023stuck} or authentication mechanisms~\cite{cai2024famos}, are similarly constrained. They enforce a static, predefined rule but remain fundamentally \textit{context unaware}, unable to distinguish a benign transaction like ordering food from a high-risk one like sending a gift to a livestreamer. Even leveraging AI models to real-time analyze screen content, though more powerful, faces a more fundamental and long-term limitation: inherent fallibility. These models suffer from incomplete knowledge~\cite{lewis2020retrieval}, are prone to hallucinations~\cite{ji2023survey}, and remain vulnerable to adversarial bypass~\cite{liu2024making,guesmi2024dap}. These problems are still unresolved and cannot be fixed by short-term engineering solutions. Consequently, any security paradigm that relies solely on autonomous detection will inevitably produce both false positives (blocking benign activities and frustrating users) and false negatives (failing to detect novel threats), and thus cannot provide reliable protection.

To address the limitations of existing automated detection approaches, we argue that \textit{alongside existing prevention and detection solutions, there is also a need for post-hoc forensic analysis framework that enables parents to investigate children's risk within their interaction history}. The key does not lie in creating better autonomous detectors, but in a post-hoc framework designed to empower human oversight. This framework can preserve a faithful record of runtime evidence from the device. It helps guardians, a child's parents or legal protectors, investigate missed threats (false negatives) by retaining captured evidence around salient interactions, and it provides the context needed to verify automated alerts and dismiss false positives. In this way, the defense paradigm shifts from pursuing an unattainable goal of perfect, autonomous prevention to enabling robust, human-in-the-loop, post-hoc forensic analysis.

Unfortunately, existing post-hoc mechanisms are unsuitable for this purpose. Relying on built-in app histories~\cite{ghosh2018safety,ghosh2018matter,beigi2019protecting} (\eg, browsing logs, payment records), while efficient, forces guardians into a labor-intensive and unreliable process of manually inspecting a fragmented patchwork of disparate logs. Furthermore, these records are often incomplete, lack the rich visual context of interactions, and can be easily deleted by the child, compromising their integrity. 
Conversely, state-of-the-art approaches like Microsoft Recall~\cite{recall} and Bytedance MineContext~\cite{minecontext}, which continuously capture screen snapshots at fixed intervals (\eg, every few seconds) to provide historical context, are unsuitable for this scenario for two fundamental reasons. First, such systems are designed for high-end desktop PCs, with prohibitive hardware requirements that are practically infeasible on resource-constrained mobile devices. Second, their core function is to save and retrieve discrete snapshots of activity, allowing a user to ``retrace your steps'' through a timeline. They are not designed to understand and summarize the semantic narrative of a user's sequential interactions~\cite{xu2025autolife,arakawa2024prism}.
\revise{These shortcomings leave a critical gap for a system that can be both semantically rich and resource-efficient. Existing work has not yet bridged this gap with an end-to-end, on-device solution.}

To bridge this gap, we address this deficiency by adapting the concept of provenance from system security to the mobile user interface. In digital forensics, provenance refers to the comprehensive history of a digital object. It documents the sequence of events leading to the current state rather than just the final outcome. This historical context allows analysts to distinguish a legitimate modification from a malicious attack~\cite{milajerdi2019holmes,li2023nodlink,king2003backtracking,han2020unicorn,jiang2023auditing}. We argue that \textit{protecting children online requires a similar paradigm}. A static image of a payment confirmation is ambiguous; however, the provenance of that page, whether the user reached it via a food delivery menu or a gift shop in a live stream, reveals the true intent.

To realize this vision, we propose \tool, \revise{the first end-to-end system that enables post-hoc, human-led child safety forensics on resource-constrained mobile devices}. Extending the principle of system provenance to the mobile domain, \tool constructs a structured, queryable record of interaction sequences, which we introduce as GUI Provenance. Unlike raw screen recordings which are unstructured and resource-heavy, \tool leverages a Multimodal Large Language Model (MLLM) to translate the temporal GUI stream into a human-understandable narrative. The system is designed to be both effective, providing the rich visual context that system logs lack, and efficient, utilizing an intelligently distilled GUI stream that captures only semantically important moments. Through this architecture, \tool empowers guardians with a natural language query interface to effectively investigate a child's digital activities.

However, it introduces three specific technical challenges that prior work has not been designed to address. 
\begin{itemize}

    \item \textbf{Efficient and Intelligent Capture (C1)}: the system must intelligently capture the salient temporal information from the high-frequency GUI stream without overwhelming the resource-constrained mobile devices. To overcome this challenge, we introduce an action-guided evidence distillation framework that intelligently reduces the overhead of sampling and the subsequent analysis. We leverage the lightweight stream of user interaction events (\eg, taps, swipes) as a marker to identify moments of high analytical value. Our system therefore invokes expensive MLLM analysis only on a small set of highly-curated visual evidence captured around these key interactions, making the entire approach both efficient and effective.
    
    \item \textbf{Semantic Interpretation (C2)}: the raw visual evidence must be accurately and interpretably translated into a semantically rich, human-understandable record. To tackle the second challenge, we design a policy-driven analysis module that employs a Chain of Thought prompting strategy, compelling the MLLM to function as a transparent reasoning engine that translates visual data into a structured record.
    
    \item \textbf{Provenance Storage and Query (C3)}: the resulting heterogeneous data, which is a mix of structured metadata, unstructured text, and visual evidence, must be efficiently stored and made accessible through a powerful query interface for human-led investigation. To overcome this challenge, we architect a unified hybrid data store and a two-stage forensic query engine, a system that supports both broad, semantic event retrieval and fine-grained, detailed event interrogation.
\end{itemize}

We implemented a prototype of \tool and rigorously evaluated it on a new, comprehensive benchmark dataset comprising 295 realistic interaction clips. This dataset was constructed through a three-stage, IRB-approved process, including formative interviews with parents to ground our 32-participant user study in real-world risk scenarios.
Our evaluation shows that \tool demonstrates high provenance fidelity, achieving a 95.23\% Macro-F1 score in labeling the risk categories of logged events, outperforming a baseline that relies on OCR and keyword matching. Furthermore, our distillation pipeline reduces the number of analyzed visual tokens by more than 89.2\% compared to a standard periodic sampling approach adopted by current state-of-the-art industrial approaches. Crucially, our two-stage forensic query engine proves highly effective, retrieving the correct event as the top result for more than 90.20\% of natural language queries and answering detailed follow-up questions with an accuracy of 88.24\%.

\revise{In summary, this paper makes the following key contributions:
\begin{itemize}
    \item We introduce \textbf{GUI Provenance}, an interaction-centric concept for post-hoc child safety forensics. It treats the GUI stream as a visual language that captures user intent and application response over time. This gives guardians queryable evidence for reviewing risky episodes, complementing automated defenses rather than replacing them.

    \item We design and implement \tool, the first end-to-end system that realizes GUI Provenance on resource-constrained mobile devices. Its novelty lies in the system-level integration of interaction-driven capture, hierarchical evidence distillation, on-device multimodal analysis, and human-in-the-loop forensic querying. The action-guided capture pipeline uses lightweight interaction events to select the frames worth analyzing, reducing the visual data sent to the MLLM by over 89\% and keeping on-device inference within mobile power and memory budgets.

    \item We evaluate \tool on a new benchmark with 295 realistic interaction clips from a 32-participant user study, including minors aged 12--18. \tool achieves 95.23\% Macro-F1 in risk logging and over 90\% Recall@1 in forensic retrieval. An end-to-end evaluation on three modern smartphones shows that the full pipeline runs on-device with measured overhead suitable for post-hoc auditing under our one-hour workload.
\end{itemize}}

\noindent \textbf{Code Availability.} %
We provide an repository (\url{https://github.com/PKU-ASAL/GUIAuditor}) containing the end-to-end pipeline (without any real user data). This allows others to rerun our procedures on synthetic or locally constructed inputs.

\section{Background}
\label{sec:background}

This section first establishes the taxonomy of online risks that defines the scope of threats we address. It then provides a brief overview of existing defense paradigms, setting the stage for our threat model and rationale in \S~\ref{sec:threat_model}.

\subsection{A Taxonomy of In-App Risks for Children}
\label{subsec:taxonomy}

As mobile phones become increasingly common among children, the online risks they face are also increasing. Recent reports~\cite{cybersafekids,european2023influence} indicate that more than 94\% of children aged 8-12 own a smart device, and this number rises to 99\% for children aged 12-14. This widespread adoption not only expands children's access to digital resources and social opportunities but also exposes them to a broader range of online risks~\cite{zhao2019make,zhang2024navigating,di2020ui,bbc-gromming,guo2024moderating}. With greater connectivity and more frequent unsupervised usage, children are increasingly susceptible to encountering harmful content, interacting with strangers, falling victim to scams, or inadvertently disclosing personal information. As a result, the likelihood of children being harmed by online risks is significantly heightened in today's digital environment.
To systematically address the diverse threats children face, we establish a risk taxonomy adapted from foundational work and tailored to the modern mobile GUI ecosystem. Our framework is informed by governmental studies~\cite{cybersafekids,european2023influence,cac-regulations} and the well-established ``4Cs'' classification (Content, Contact, Conduct, Commerce) by Livingstone~\cite{livingstone2009eu}.

While the ``4Cs'' provide an essential foundation, they were originally formulated in the context of the open web. The shift to the modern mobile app ecosystem necessitates specific adaptations to this framework for two primary reasons. First, the friction-less nature of mobile payments has transformed broad ``Commerce'' risks into high-frequency, deceptive ``Financial'' traps (\eg, loot boxes, streamer gifting) that are deeply embedded in legitimate applications~\cite{aagaard2022game,wu2023malicious}. Second, the data-centric architecture of mobile devices introduces unique ``Privacy'' vectors—such as malicious permission requests for location or camera access, which were less central in the desktop era~\cite{cao2021large,zhang2023android}.

\revise{We therefore adapt the classic ``4Cs'' model into four domains for mobile in-app risks. The taxonomy maps directly to the attack vectors in our threat model (\S~\ref{sec:threat_model}). It keeps Content and Contact risks from the original framework, while separating Financial and Privacy risks as first-class categories to better capture the mobile-specific risks discussed above:}

\begin{itemize}
    \item \textbf{Content Risks}: Harms arising from exposure to age-inappropriate or malicious information. This aligns directly with the ``Content'' category and includes risks like encountering violent imagery on social media or extremist propaganda.
    \item \textbf{Contact Risks}: Harms arising from interaction with other individuals online. This category retains the ``Contact'' risk (\eg, sexual exploitation) and also encapsulates the interactive aspects of ``Conduct'' risk (\eg, cyberbullying), as both manifest on the GUI as harmful interactions.
    \item \textbf{Financial Risks}: Threats leading to monetary loss or fostering unhealthy consumer habits. This category is a necessary modernization of the ``Commerce'' risk. While ``Commerce'' is broad, ``Financial'' more precisely captures the specific, high-frequency risks of the mobile ecosystem, such as deceptive in-app purchases, gambling-like loot boxes, or high-value donations to livestreamers.
    \item \textbf{Privacy Risks}: Harms related to the improper collection, use, or disclosure of a child's personal data. We elevate this to a first-class category, as the data-centric nature of mobile apps presents critical privacy threats (\eg, inadvertent PII disclosure, malicious permission granting) that were less central when the ``4Cs'' framework was first formulated.
\end{itemize}

\subsection{Overview of Existing Defense Paradigms}
\label{subsec:existing_defenses}

Current technical approaches to child safety primarily fall into two categories, both focused on automated, real-time risk prevention or detection.

First, \textbf{rule-based controls} operate on predefined policies. These include coarse-grained mechanisms like OS-level app blockers~\cite{apple-screen-time,google-family-link} and network filters~\cite{adguard}, as well as system-level controls such as OS permission managers~\cite{alsoubai2022permission,tahaei2023stuck} or authentication mechanisms~\cite{cai2024famos}. These tools are designed to enforce static boundaries (\eg,``block this app'' or ``deny this permission'').

Second, \textbf{AI-based analysis} represents a more advanced approach that attempts to analyze on-screen content in real-time to identify risks~\cite{recall,recall-block}.

While these paradigms provide a crucial first line of defense, they suffer from a fundamental limitation: they are context-unaware.

\noindent\textbf{The Context Gap.} Automated systems analyze events in isolation. Rule-based blockers enforce static boundaries (\eg, ``block App X''), while real-time AI models scan individual screen frames for prohibited content. However, the risks we target are often not contained in a single screen or event but are embedded in the temporal sequence of interactions.

\begin{figure}[h]
    \centering
    \includegraphics[width=0.7\columnwidth]{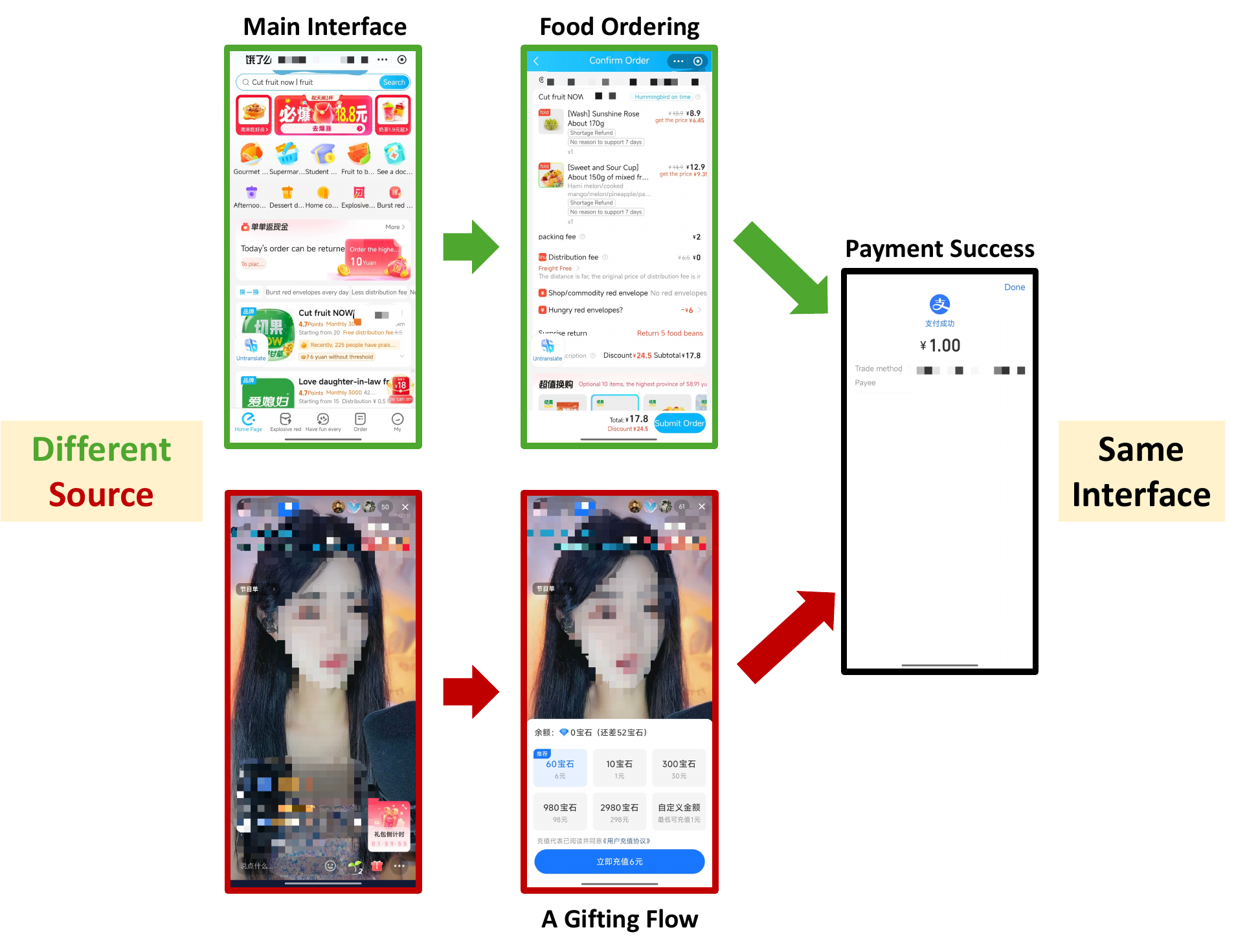}
    \caption{A motivating example. Upper: An interaction flow of a child ordering food on a food delivery app. Lower: An interaction flow of a child gifting to a livestreamer on a social platform. The two flows are both from mini-apps inside the same super app, and they both reach the same screen of a payment confirmation, but the child's intent is different.}
    \label{fig:motivating}
\end{figure}

\noindent\textbf{Motivating Example.} As illustrated in \autoref{fig:motivating}, a final payment confirmation screen is visually identical for a benign activity (\eg, ordering food) and a high-risk one (\eg, gifting to a livestreamer). An automated, real-time system cannot distinguish between them without the historical context. It is forced to choose between a false negative (allowing the risky gift) or a false positive (blocking the benign food order).

This deficiency is particularly acute in the era of super apps, where a single application (\eg, Wechat or TikTok) serves as a portal to a multitude of diverse functionalities, ranging from messaging to shopping and gaming~\cite{cai2024famos,cai2025can,wang2023uncovering}. This context gap renders purely autonomous prevention insufficient, necessitating a paradigm shift towards context-aware forensics.

\subsection{From System Provenance to GUI Provenance}
\label{subsec:provenance}

\noindent\textbf{The Concept of Provenance.} In system security, provenance refers to the comprehensive history of a digital object, documenting the sequence of events and causal relationships that led to its current state. Traditionally, provenance analysis operates at the OS level to backtrack intrusions and identify malicious activity~\cite{milajerdi2019holmes,li2023nodlink,king2003backtracking,han2020unicorn,jiang2023auditing}. The core insight is that understanding how a state was reached is often more critical than the state itself.

\noindent\textbf{Rationale for a Post-hoc Paradigm.} We argue that applying this insight to child safety is necessary because current automated defenses fail to capture context (as discussed in \S~\ref{subsec:existing_defenses}). Instead of pursuing the unattainable goal of perfect, autonomous prevention, we propose a complementary, post-hoc forensic paradigm. The defensive goal shifts from blocking real-time events to preserving faithful evidence of the interaction sequence. This approach mitigates the risk of false negatives by retaining captured evidence around salient interactions, allowing a human decision-maker to distinguish benign behavior from true harm based on the available context.

\noindent\textbf{Defining GUI Provenance.} To materialize this paradigm, we introduce GUI Provenance. Unlike system provenance which tracks opaque data flows, GUI Provenance tracks the semantic flow of user interactions. We define it as a structured, queryable record that captures the semantic essence of a user's interaction sequence (\eg, distinguishing a ``Food Mini-App'' flow from a ``Live Stream Gifting'' flow). This concept serves as the theoretical foundation for our system design. This approach directly mitigates the risk of false negatives by enabling the guardian (the human-in-the-loop) to later review the captured flow (\eg, ``Food Mini-App $\rightarrow$ Checkout'' vs. ``Live Stream Mini-App $\rightarrow$ Gifting $\rightarrow$ Checkout''), providing the necessary context to distinguish benign behavior from true harm.

\section{Threat Model and Rationale}
\label{sec:threat_model}

In this section, we define the threat model, the involved actors, and crucially, the fundamental limitations of existing defenses that motivate our post-hoc forensic paradigm.

\subsection{Operational Scenario and Attacker Model}
\label{subsec:scenario_attacker}

The operational scenario involves a child using a standard, unmodified smart device (\eg, an Android smartphone). We assume the device runs a non-compromised OS with legitimate applications installed from official marketplaces~\cite{cai2025can,liu2025hijacking}. The child, as a non-expert user, may be susceptible to risks like social engineering or deceptive UI patterns~\cite{di2020ui}.

\noindent \textbf{Attackers}. The attacker represents a spectrum of threats whose attack surface is the GUI content of these legitimate applications. We assume attackers operate through in-app channels and cannot compromise the underlying OS~\cite{cai2024famos,cai2025can}. Aligned with our risk taxonomy (\S~\ref{subsec:taxonomy}), we categorize attackers by their nature:
\begin{itemize}
    \item \textbf{Contact \& Privacy Attackers}: Malicious human actors (\eg, groomers, scammers) who interact with the child via legitimate communication channels (\eg, in-game chat, social media direct messages) to elicit private information or build deceptive trust.
    \item \textbf{Financial Attackers}: Platforms or developers whose apps use mechanisms (\eg, in-app purchases, loot boxes, streamer-gifting) that can lead to significant, unauthorized, or unhealthy financial transactions.
    \item \textbf{Content Attackers}: Algorithmic systems (\eg, recommendation engines) or content creators that push age-inappropriate, harmful, or radicalizing material to the child.
\end{itemize}

\subsection{Defender Model}
\label{subsec:defender}

\noindent\textbf{The Defender.} We define the defender as the child's parent or legal guardian. The defender is characterized by two key constraints: 1. \textbf{Remote Supervision:} The defender is typically physically remote from the child during device usage and lacks real-time visibility into the screen. 2. \textbf{Non-Expert Analysis:} Unlike professional security analysts, the defender is likely a non-expert user who requires intuitive, high-level tools to interpret technical risks.

\noindent\textbf{Role in the Loop.} In our post-hoc paradigm, the defender acts as the final ``human-in-the-loop'' decision maker. \tool serves as their technical instrument, bridging the gap between raw device data and their semantic understanding of safety.

\subsection{System Scope and Assumptions}
\label{subsec:scope}

Based on this rationale, we define the scope and assumptions for \tool:
\begin{enumerate}
    \item \tool is installed with guardian and child consent and runs with necessary OS permissions.
    \item All data capture (GUI events) and analysis (MLLM inference) are performed exclusively on-device to protect privacy. No screen content is transmitted to external servers.
    \item \tool is designed as a decision-support system that provides timely, context-rich insights, empowering the guardian to take appropriate action, rather than acting as an autonomous detection system.
    \item We explicitly scope our work to interaction-mediated risks, which require a user action (\eg, a tap, swipe, or text entry) to manifest. A purely passive risk (\eg, receiving a malicious image but never opening or interacting with it) is outside our current event-driven capture model. However, any subsequent user interaction with that content (\eg, tapping to view it, replying) would be captured and logged as part of the provenance record.
\end{enumerate}

\section{Design of \tool}

\begin{figure*}[t]
    \centering
    \includegraphics[width=1\textwidth]{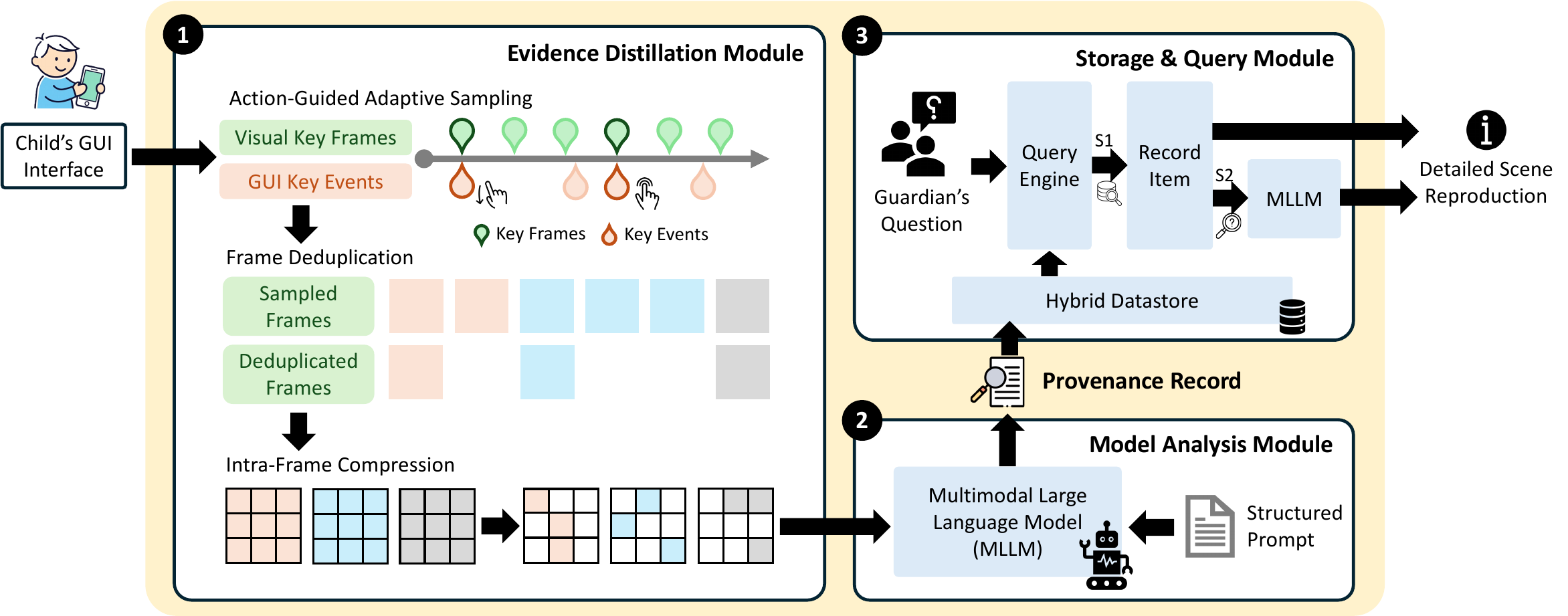}
    \caption{The architecture of \tool, illustrating the end-to-end workflow from evidence capture to human-led forensic investigation. The Evidence Distillation Module (\compreduce) ingests the raw stream and applies a three-step hierarchical process to produce a compact set of core visual evidence. The Model Analysis Module (\companalysis) leverages a MLLM guided by a structured, policy-driven prompt to translate this evidence into a structured provenance record, which is persisted in a hybrid database. The Storage and Query Module (\compstorage) enables guardian-led investigation via a two-stage query engine. It first retrieves relevant event items from the database based on the guardian's question and then allows for fine-grained interrogation of a specific event's visual evidence by re-invoking the MLLM.}
    \label{fig:method}
\end{figure*}

Our work is guided by the insight that a post-hoc, human-in-the-loop forensic paradigm is necessary to overcome the limitations of automated detection. This section details the design of \tool, the end-to-end framework we built to materialize this new paradigm. Our design is architected to overcome the three technical challenges we identified in \S~\ref{sec:intro}. Specifically, \tool is composed of three primary, collaboratively designed modules, with each module purpose-built to address one of the challenges. The Evidence Distillation Module (\compreduce) is designed to overcome C1. The Model Analysis Module (\companalysis) is designed to overcome C2. Finally, the Storage and Query Module (\compstorage) is designed to overcome C3. The overall architecture, illustrating how these modules work together, is shown in \autoref{fig:method}.

\subsection{Evidence Distillation Module (\compreduce)}
\label{subsec:action-guided-evidence-distillation}
The primary goal of \tool is to create a concise, yet semantically complete, body of evidence for further analysis. The Evidence Distillation Module is designed to solve the foundational challenge of capturing evidence from raw, large-scale GUI interactions in a manner that is both computationally efficient and semantically precise. Traditional video reduction techniques~\cite{agarwala2004keyframe,asha2018key,zhang2024universal,nair2022r3m} are ineffective for this purpose because the nature of GUI interaction streams presents unique characteristics. We identify three core characteristics in GUI scenarios: event-driven semantics, inter-frame redundancy, and intra-frame sparsity.

First, GUI interactions exhibit \textit{event-driven semantics}. Unlike in natural videos, the moments of semantic importance are anchored to discrete user actions, rather than to the magnitude of visual change. Second, they suffer from high \textit{inter-frame redundancy} due to asynchronous UI updates. After a user action, the screen may remain static or update slowly, generating numerous visually identical frames. Third, GUIs are characterized by \textit{intra-frame sparsity}, where a single frame is spatially sparse, with most of its area consisting of static, low-information elements like backgrounds or toolbars.

In considering these characteristics, we implement a three-step hierarchical evidence distillation pipeline that progressively reduces the raw stream into a compact set of core visual evidence before invoking costly MLLM analysis.

\subsubsection{Step 1: Action-Guided Adaptive Sampling} 
To address the challenge of event-driven semantics, our approach is built upon the insight that user interaction events (\eg, taps, swipes) are a high-fidelity proxy for user intent and semantically important moments. Standard video summarization methods that rely on visual change are ineffective, as they would systematically miss critical but visually static events, such as a single tap on a ``Confirm Payment'' button. 
We select user interaction events as the core trigger because the vast majority of online harms ultimately require a user's interactive action to be completed. Whether it is clicking ``Confirm Payment'' for a Financial Risk, sending a message in a Contact Risk scenario, or tapping ``Allow'' to grant sensitive permissions for a Privacy Risk, these interactions are the critical execution points in the risk chain.

To leverage our insight, we use \textit{Action-Guided Adaptive Sampling}. Instead of continuous monitoring, the system operates in a lightweight, near-zero-overhead standby state. It activates only when a \textit{GUI Key Event} (a tap or swipe) is detected. At that point, it captures a temporal window of frames around the event at a predefined dense frequency, $F_{high}$ (\eg, 2 fps), for $t_{window}$ seconds. This mechanism yields a candidate set of frames around interactions, significantly reducing the volume of data requiring further analysis.

\revise{We choose $F_{high}=2$ fps and $t_{window}=2$ seconds based on two observations. First, most mobile GUI state transitions, such as page loads, dialog appearances, and content rendering, stabilize within 1--2 seconds after a user action~\cite{yan2012fast}. Second, users typically leave at least 500\,ms between consecutive actions~\cite{cai2025can,jesdabodi2015understanding,nishida2023single}, which matches the sampling period of $1/F_{high}=500$\,ms. Thus, a 2-second window at 2 fps yields roughly four candidate frames, enough to cover both the transient state after an action and the stable post-transition state. Frames that capture intermediate animations or loading screens are later removed by pixel-level deduplication (Step 2) and semantic compression (Step 3). We validate this choice through the sensitivity analysis in \S~\ref{subsec:rq1}.}

\revise{\noindent\textbf{Pre-action Context.} One concern is whether the system preserves enough context from \textit{before} the user's action. In mobile GUIs, the relevant context is usually visible \textit{when the action is performed}. If a child taps a notification, the first captured frame contains the notification content. If a child clicks ``Confirm Payment,'' the item name, price, and merchant are shown on the same screen at the time of the click. For direct-manipulation interfaces, the trigger and its context typically appear in the same frame. The frame captured at the GUI Key Event therefore serves as the pre-action context, while the following frames within $t_{window}$ capture what happens next.}

\revise{\noindent\textbf{Handling of Continuous Interaction.} When a user performs multiple actions in rapid succession (\ie, with inter-action intervals shorter than $t_{window}$), the capture windows naturally overlap. In our implementation, overlapping windows are merged into a single extended capture session: the system stays in the active capture state as long as new events keep arriving within $t_{window}$ of the previous event, and returns to standby only after $t_{window}$ elapses with no new event. This merging keeps interaction-relevant frames from rapid bursts in the same capture session, rather than treating each action as an isolated window. During prolonged high-frequency interaction (\eg, rapid repetitive tapping in a game), the downstream deduplication stages (Steps 2 and 3) act as natural throttles, filtering out the resulting near-identical frames. In practice, even under sustained interaction, the number of semantically distinct frames passed to the MLLM stays small, as our token reduction results in \S~\ref{subsec:rq1} show.}

\subsubsection{Step 2: Fast Deduplication via Pixel Similarity} 
To tackle the high inter-frame redundancy caused by UI update delays and repetitive user actions, our second distillation step employs a lightweight pre-filter. A naive approach would be to retain all frames captured in the previous step, but this would lead to significant redundant processing, such as when a child rapidly taps the screen to collect rewards in a game. To eliminate this redundancy efficiently, we introduce a fast, low-cost deduplication filter based on pixel similarity. For the candidate set of frames $\{f_1, f_2, \ldots, f_n\}$ generated in the previous step, we compute the L1 distance between the pixel values of adjacent frames, defined as:

\begin{equation}
    d(f_i, f_{i+1}) = \frac{1}{3M} \sum_{j=1}^{M}\sum_{k=1}^{3} |p_{i,j,k} - p_{i+1,j,k}|
\end{equation}
where $M$ is the total number of pixels in a frame, and $p_{i,j,k}$ is the pixel value of the $k$-th channel (RGB) at the $j$-th position in frame $f_i$. If the distance $d(f_i, f_{i+1})$ is below a predefined threshold $\tau_{pixel}$, indicating the frames are nearly identical, the latter frame $f_{i+1}$ is discarded. This step efficiently removes exact or near-exact duplicates produced by rapid, repetitive interactions, ensuring that only visually distinct frames proceed to the final, most resource-intensive stage of analysis.

\subsubsection{Step 3: Semantic Intra-frame Compression}
Finally, to address the intra-frame sparsity of GUIs, our last distillation step performs patch-level semantic diffing to isolate only the visually changed and semantically meaningful regions within a frame. Even a single keyframe is spatially sparse, and naively tokenizing the entire high-resolution image is inefficient~\cite{lin2025showui,wang2025dymu}.
To address this, we modify recent works on video token reduction~\cite{fu2024framefusion,sun2025llava,yao2025timechat,jin2024chat} to adapt it for GUI interactions. Specifically, each candidate frame $f_i$ is first partitioned into a grid of non-overlapping patches. A lightweight Vision Transformer (ViT) encoder then maps each patch at a given spatial coordinate $(h,w)$ to a feature-level vector, $v_{i,(h,w)} \in \mathbb{R}^{d}$. To identify temporal redundancy, we then compare the feature vectors of spatially-aligned patches between two consecutive frames, $f_i$ and $f_{i+1}$, by calculating their cosine similarity:

\begin{equation}
    C_{i,i+1, (h,w)} = \frac{v_{i,(h,w)} \cdot v_{i+1,(h,w)}}{\|v_{i,(h,w)}\| \|v_{i+1,(h,w)}\|}
\end{equation}

A token $v_{i+1,(h,w)}$ is considered redundant if its similarity to its predecessor $v_{i,(h,w)}$ exceeds a threshold $\tau_{feat}$. We discard the redundant tokens and keep the non-redundant ones. By computing the cosine similarity of feature vectors for spatially-aligned patches between consecutive keyframes, we can identify and discard patches that have not changed semantically. This process effectively removes static regions like backgrounds and toolbars, producing a final, highly compact set of core visual tokens that represent the semantic changes on screen for the MLLM to analyze.

\subsection{Model Analysis Module (\companalysis)}
Once \compreduce has produced a compact set of core visual evidence, \companalysis is responsible for performing high-level semantic interpretation. This module takes the distilled visual tokens as its primary input and is designed to produce a structured JSON object as its output. This JSON object, containing a natural language summary, a risk classification, and a detailed rationale, forms the core of our provenance record. It is then passed to \compstorage to be persisted, enabling a guardian to later query the event using natural language.

To achieve this translation, the module is powered by a Multimodal Large Language Model (MLLM). However, a naive approach that treats the MLLM as a black-box classifier would lack the explainability that is crucial for a forensic system. Therefore, to address our second design goal of accurate semantic interpretation, we architect this module as a transparent, policy-driven reasoning engine. We implement this by constraining the MLLM's behavior using a sophisticated structured prompt that embodies a security policy and employs a Chain-of-Thought (CoT) approach, which instructs the model to externalize its reasoning process before delivering a final verdict. The prompt consists of three key dimensions:

\begin{itemize}[leftmargin=2em]
    \item \textbf{Role Definition}: The MLLM is instructed to act as a digital child safety analyst, whose primary goal is to audit GUI interactions for potential harm to minors.
    
    \item \textbf{Policy Definition}: Our risk taxonomy (Content, Contact, Financial, Privacy) is presented to the model not merely as classification labels, but as a formal security policy. Each category is accompanied by a detailed description and examples, defining the rules against which the interaction sequence must be judged.
    
    \item \textbf{Structured Reasoning and Record Generation}: We prompt the model to carry a multi-step reasoning process to analyze the visual evidence before generating the final structured JSON output. This ensures the final output is a direct product of an explicit analytical process. The generated record contains several key fields:
    \begin{itemize}[label=-]
        \item \texttt{abstractive\_summary}: The model is instructed to provide a concise, natural-language description of the event first. This serves as a human-readable summary for the guardian, offering immediate context without needing to review the raw evidence.
        \item \texttt{risk\_type}: Following its reasoning, the model must classify the event into one of the categories from the defined policy (or a No Risk category). This structured label is crucial for high-level alerting and efficient, metadata-based filtering.
        \item \texttt{risk\_description}: Crucially for explainability, the model must provide a specific, detailed rationale explaining why the event was flagged with a particular risk type, linking its conclusion back to the policy. The rationale should be grounded in the visual evidence and the policy.
    \end{itemize}
\end{itemize}

Following the generation of this complete JSON record, a handling process is initiated. If the \texttt{risk\_type} is determined to be No Risk, the record is discarded. Otherwise, if a potential risk is identified, the record is passed to the \compstorage for the following investigation.

\subsection{Storage and Query Module (\compstorage)}
The logging of a significant event is often not the endpoint, but the beginning of a necessary forensic analysis by the guardian. This module is the centerpiece of our human-in-the-loop paradigm, designed to empower the guardian with a powerful yet intuitive interface for investigation. Building such a module, however, presents two central challenges. First is the \textit{heterogeneous data challenge}: the provenance records generated by \companalysis are complex, composite objects containing a mix of structured metadata, unstructured text, and visual evidence, which are difficult to store and query cohesively. A traditional relational database is ill-equipped for the semantic text search required by our system, while a simple file-based log would make any form of efficient querying impossible. Second is the \textit{multi-granularity query challenge}: a real human investigation is a multi-step process that moves from broad, exploratory questions to specific, detailed ones, and the query system must naturally support this workflow. To address these challenges, our module consists of two core components.

\subsubsection{Unified Hybrid Data Store}
To address the challenge of heterogeneous data, we designed a Unified Hybrid Data Store to efficiently manage our composite GUI Provenance records. Our architecture is built upon a modern relational database system that is extensible with vector support, allowing all components of a record to be managed within a single, cohesive system. Each provenance record corresponds to a single row in our database and is composed of the following key fields:

\begin{itemize}%
    \item \textbf{Structured Metadata}: This includes fields such as a unique \texttt{event\_id}, a \texttt{timestamp}, and the \texttt{risk\_type} label generated by \companalysis. These fields are indexed for fast, exact-match filtering.
    
    \item \textbf{Unstructured Text}: This includes the \texttt{abstractive\_summary} and \texttt{risk\_description} texts generated by \companalysis, providing a human-readable narrative. 
    
    \item \textbf{Visual Evidence}: To balance privacy and the need for scene reproduction, we do not store the original raw video or continuous screen recordings. Instead, we only store the high-level core visual tokens generated by \compreduce. This intermediate representation provides sufficient data for detailed follow-up queries while minimizing the privacy-sensitive data stored on the device.
    
    \item \textbf{Summary Embedding}: To enable semantic search, we generate a high-dimensional \texttt{summary\_embedding} vector. Specifically, we concatenate the unstructured text fields and pass them to a dedicated text embedding model. We deliberately chose to embed the MLLM-generated textual narrative rather than the visual tokens (via CLIP~\cite{radford2021learning}, which is designed for natural images). This is because our events are inherently sequential and domain-specific (GUIs), making the textual a more direct and suitable representation for semantic matching with a guardian's natural language query. This vector is then stored alongside the other data in the same row.
\end{itemize}

\subsubsection{A Two-Stage Forensic Query Engine} 
To address the multi-granularity query challenge, our query engine is designed to mirror a guardian's natural workflow. A typical investigation is not a single query but a multi-step process that begins with broad exploration to identify events of interest, followed by a deep-dive interrogation to extract specific details from the evidence. To directly support this, we designed a novel Two-Stage Forensic Query Engine built upon our hybrid datastore.

\noindent\textbf{Stage 1: Event Retrieval}. This stage is designed to support the initial, broad exploration phase of an investigation. When a guardian poses a general question, the engine's goal is to locate the most relevant provenance records. It first analyzes the natural language query to determine the optimal retrieval strategy. 
For questions with structured entities, such as dates or explicit risk categories (\eg, ``What financial risks occurred yesterday?''), the engine constructs a precise SQL clause for efficient, exact-match filtering. For more abstract, semantic questions (\eg, ``Did my child encounter any bullying?''), it leverages vector similarity search against the stored \texttt{summary\_embedding} vectors. The unified nature of our datastore is critical here, as it allows these two retrieval mechanisms to be seamlessly combined. 

\noindent\textbf{Stage 2: Event Interrogation}. This stage supports the subsequent deep-dive analysis. Once a specific event of interest has been retrieved, the guardian can investigate it with a detailed follow-up question (\eg, ``What was the exact amount of this purchase?''). For these queries, the engine retrieves the stored visual evidence for that event and re-invokes the MLLM. It provides the MLLM with both the visual evidence and the new question, performing a fine-grained visual question answering (VQA) task to extract the specific details. This two-stage process empowers the guardian to seamlessly navigate from a high-level overview to a deep-dive analysis of specific evidence.

\section{Evaluation}
\label{sec:evaluation}
In this section, we present a comprehensive evaluation of \tool to answer the following research questions:
\begin{tcolorbox}[size=small]
\begin{itemize}
    \item \textbf{RQ1 (Provenance Fidelity):} How faithfully does \tool log and label risk episodes such that later forensic questions can be answered?
    
    \item \textbf{RQ2 (Forensic Capability):} How accurately does the forensic query engine retrieve the right events and answer follow-up factual questions?
    
    \item \textbf{RQ3 (Performance Overhead):} What is the \revise{end-to-end} on-device performance overhead (\eg, power consumption, \revise{inference latency}) of \tool, and how does it compare to a naive, continuous analysis approach?
\end{itemize}
\end{tcolorbox}

\subsection{Experimental Setup}
\subsubsection{Data Construction}
\label{subsubsec:data_construction}
A significant challenge in evaluating GUI-based child safety systems is the absence of publicly available, realistic datasets. Collecting and sharing screen recordings of real children's device usage would pose inevitable ethical and privacy risks~\cite{teng2024tool,wang2024koala}. To overcome the ethical and privacy risks while ensuring our evaluation is comprehensive and grounded in real-world risks, we constructed a new benchmark dataset through a rigorous, three-stage process.

\noindent\textbf{Stage 1: Formative Study to Identify Parental Concerns.}
To ground our scenarios in authentic parental concerns, we conducted a formative qualitative study focused on ages 10-16. This age range was selected as it represents the critical transition period into independent device usage. According to recent Ofcom reports, 96\% of children aged 5-15 are online, and crucially, more than half of ten-year-olds already own their own smartphones or tablets~\cite{ofcom2024adults,wang2021protection}. This widespread ownership exposes this demographic to heightened interaction-based risks (\eg, grooming, scams) while they still lack full adult judgment.~\cite{anderson2023teens,gommerman2024comparative}. We interviewed 8 parents to understand their primary anxieties and lived experiences regarding their children's smart device usage. These interviews were transcribed and thematically analyzed, yielding a seed set of 22 distinct risk themes and concrete examples (\eg, ``my child was tricked into buying game currency'', ``I worry about who they are talking to in that game''). This step provided the qualitative ground-truth that informed our scenario design.

\noindent\textbf{Stage 2: Systematic Risk Scenario Construction.}
\revise{Next, we constructed the full corpus of interaction scenarios. Rather than writing scripts from scratch based only on our own assumptions, we derived them from three sources:
\begin{enumerate}%
    \item \textbf{Parental Concerns:} The 22 themes identified in Stage 1.
    \item \textbf{Public Reports:} Real-world case studies of harm documented in governmental reports~\cite{cac-regulations,european2023influence}, news articles (\eg, BBC reports~\cite{bbc-gromming}), and child safety NGO findings~\cite{livingstone20214cs,cybersafekids}.
    \item \textbf{Taxonomy Coverage:} Our risk taxonomy (\S~\ref{subsec:taxonomy}), which required scenarios covering all four risk categories (Content, Contact, Financial, Privacy) and a benign ``No Risk'' category.
\end{enumerate}

Two researchers synthesized these sources to write detailed scripts for 320 interaction tasks. Each script precisely defined a user goal (\eg, ``purchase a gift for a livestreamer''), the target application, and the sequence of steps needed to complete the interaction. The resulting 320 task scripts covered a wide range of popular applications, including super-apps, standalone social media apps (\eg, TikTok, Instagram), and various mobile games, ensuring our benchmark was not limited to a single app ecosystem.} The example scripts can be seen in \autoref{tab:scenario_examples}.

\begin{table*}[t]
\centering
\caption{Illustrative examples of the task scripts from our benchmark dataset, corresponding to the five categories defined in our taxonomy (\S~\ref{subsec:taxonomy}).}
\label{tab:scenario_examples}
\begin{tabularx}{1.0\textwidth}{@{}lX@{}}
\toprule
\textbf{Category} & \textbf{Example Task Script} \\ \midrule
\textbf{Content} & Open X App, scroll through the recommendation feed until you encounter a post with violent/gory/adult imagery, and watch it for at least 5 seconds.\\ \addlinespace
\textbf{Contact} & Enter Wechat, notice a message preview from a contact containing harassing or explicit words. Open the conversation and send a reply.\\ \addlinespace
\textbf{Financial} & Enter TikTok, join a stream, find the ``Gift'' button. Purchase a ``Virtual Rocket'' and send it to the streamer. \\ \addlinespace
\textbf{Privacy} & Download a photo edit app from the app store. On first launch, accept all permission requests it asks for, including access to your Photos, Contacts, and Location. \\ \addlinespace
\textbf{No Risk} & Open Weather App, search for the weather in Tokyo, and check the forecast for the next three days. \\ \bottomrule
\end{tabularx}%
\end{table*}

\noindent\textbf{Stage 3: Data Collection and Ground Truth Labeling.}
\revise{For data collection, we recruited 32 volunteers: 16 \textit{Adult Proxies} aged 18--30 and 16 \textit{Underage Participants} aged 12--18, including 4 aged 12--13, 2 aged 14--15, and 10 aged 16--18. Adult proxies have app usage patterns reasonably close to older teenagers~\cite{ofcom2024adults,anderson2023teens} and allowed us to test high-risk scenarios (\eg, simulated violent content, financial scams) without exposing younger minors to harmful material. However, an adult-only study could miss differences in adolescent behavior, such as faster scrolling, higher touch frequency, and more rapid task switching~\cite{ofcom2024adults,anderson2023teens}. We therefore included minors across this age range to check whether our evidence distillation pipeline (\compreduce, \S~\ref{subsec:action-guided-evidence-distillation}) remains effective for younger users. All underage participation followed our IRB-approved protocol (see Ethics Considerations below, \S~\ref{subsubsec:ethics}).}

\revise{Specifically, each participant was assigned 10 tasks selected from the scenario pool under an IRB-approved risk-minimization protocol. Task assignment was stratified to keep the final dataset balanced across the five categories, while tasks involving explicit harmful visual content were assigned only to adult proxies. In total, we collected $32 \times 10 = 320$ screen-recorded interaction clips. All clips were originally recorded at 24 fps, a standard rate for most natural videos~\cite{madhusudana2021subjective}. To establish a high-quality ground truth, each collected clip was independently classified by a panel of three researchers (annotators) based only on the screen recording, without access to the task script. The annotators were tasked with identifying any risky events within the clip and assigning a risk category from our taxonomy. In cases of disagreement among the annotators, they were required to engage in a discussion to reach a consensus. To ensure the unambiguity and high quality of our ground truth, any clips for which a clear consensus could not be reached were discarded from the final dataset. After filtering, we retained 295 clips, each with an average duration of 17.2 seconds. The final distribution of labeled events across the five categories is presented in \autoref{tab:ground-truth}. As shown, the dataset was deliberately constructed to be reasonably balanced, ensuring that our evaluation is not biased towards any single type of threat.
}

\revise{\subsubsection{Ethics Considerations}
\label{subsubsec:ethics}

This study obtained the Institutional Review Board (IRB) approval from our institution. All procedures followed the approved protocol and IMWUT's ethics policy. 

Because this research involves minors and safety-sensitive mobile interactions, we designed the study to prioritize risk minimization, informed participation, and strong privacy protections. The collected interaction data were used solely to validate our safety auditing pipeline.

\noindent\textbf{Informed Consent.} We obtained informed consent from all participants, with an additional dual-consent procedure for minors. For the formative interviews (Stage 1), all parental participants provided written informed consent, and their responses were anonymized before analysis. For data collection (Stage 3), all 16 adult participants signed written informed consent forms. For the 16 minor participants (aged 12--18), we obtained written informed consent from their legal guardians \textit{and} written assent from the minors themselves. All minor participants were aged 12 or above, well beyond the age of 7 at which children are generally considered capable of providing meaningful assent in research~\cite{ucsf_irb_children}. Thus, all minor participants were able to understand the study purpose and voluntarily agree to participate.

\noindent\textbf{Participant Safeguards.} We followed the IRB-approved risk-minimization protocol during data collection, with additional safeguards for minors. First, guardians were present throughout each session and could stop participation at any time. Second, task assignment followed a risk-tiered protocol: tasks involving explicitly harmful visual content (\eg, graphic violence or sexual material) were assigned only to adult proxies, so no minor was directly exposed to explicit harmful material during the study. This protocol allowed the dataset to cover high-risk scenarios while minimizing risk to underage participants.

\noindent\textbf{Data Protection.} All collected data are stored with strict access controls and will not be publicly released. Participants were informed that screen recordings would be used only for validating the auditing pipeline. The data were stored on encrypted, access-controlled institutional servers, and only researchers listed on the IRB protocol could access them. Raw screen recordings were retained only for the duration of the validation process and will be deleted upon study completion.}

\begin{table}[h]
    \centering
    \caption{\revise{The distribution of labeled events across our five categories (the four risk types plus a benign No Risk category).}}
    \label{tab:ground-truth}
    \begin{tabular}{lccccc}
        \toprule
        \textbf{Category} & \textbf{Content} & \textbf{Contact} & \textbf{Financial} & \textbf{Privacy} & \textbf{No Risk} \\
        \midrule
        Count      & 61 & 57 & 59 & 56 & 62 \\
        Percentage & 20.68\% & 19.32\% & 20.00\% & 18.98\% & 21.02\% \\
        \bottomrule
    \end{tabular}
\end{table}

\subsubsection{Implementation Details and Baselines} 
\label{subsubsec:setup}
Our experimental setup is designed to separately evaluate the effectiveness and the performance overhead of our system.

\revise{
\noindent \textbf{Evaluation Platforms.} For the effectiveness evaluation, we deployed a prototype of \tool's analysis pipeline on a server equipped with an Intel Xeon Platinum 8358 CPU and an NVIDIA RTX A6000 GPU. This server setup gives all methods the same runtime environment on the collected clips. \revise{It also isolates provenance fidelity and query accuracy from device-level resource variability. We separately evaluate deployment cost by measuring the \emph{end-to-end} on-device overhead of the full \tool pipeline, from event detection to on-device MLLM inference, on three modern smartphones: a Xiaomi 15 (Snapdragon 8 Elite, 12\,GB RAM, 5,240\,mAh), a Google Pixel 9 Pro (Tensor G4, 16\,GB RAM, 5,060\,mAh), and a Vivo X200 Pro (Dimensity 9400, 16\,GB RAM, 6,000\,mAh). Testing across multiple devices reduces the chance that the results reflect one SoC vendor's implementation.}}

\noindent\textbf{Model and Hyperparameters.} \revise{Our framework is designed to be model-agnostic. For our experiments, we selected a state-of-the-art Multimodal Large Language Model, Qwen2.5-VL-3B, as our core analysis engine. A model of this size (3B parameters) represents a reasonable balance between high performance and practical on-device deployment~\cite{lu2025bluelm,xu2025fast,yao2024minicpm}. For on-device inference (RQ3), we use the MLC-LLM mobile runtime~\cite{mlc-llm} with its q4f16\_1 4-bit quantization scheme, with the full overhead analysis presented in \S~\ref{subsec:rq3_overhead}}. Key hyperparameters for \compreduce were set as follows: the adaptive sampling frequency $F_{high}$ was set to 2 fps within a $t_{window}$ of 2 seconds around an event. The pixel similarity threshold $\tau_{pixel}$ for fast deduplication was set to 30, and the feature-level token similarity threshold $\tau_{feat}$ was set to 0.5.

\noindent\textbf{Baseline Methods.} As no existing parental control systems (\eg, Apple's Screen Time) target the post-hoc, fine-grained and intra-application level of our work, a direct comparison is not feasible. We therefore design and implement several baseline systems to serve as intuitive approaches for a rigorous comparison, including a naive method based on Optical Character Recognition (OCR). The full suite of baselines is as follows:

\begin{itemize}%
    \item \toolnaive (Naive Continuous Analysis): This baseline represents the most direct, brute-force visual analysis approach. It involves attempting to feed a continuous, high-framerate screen recording directly to the MLLM for analysis. It serves as a theoretical upper bound for accuracy but is expected to be practically infeasible due to token limits and prohibitive resource costs.
    
    \item \tooluniform (Periodic Snapshot / Industry Standard): This baseline represents the prevailing industrial paradigm for user history retrieval, exemplified by desktop-grade solutions like Microsoft Recall~\cite{recall} and ByteDance MineContext~\cite{minecontext}. However, these systems rely on uploading screenshots to cloud-based VLM APIs for reasoning, which poses severe privacy risks for sensitive data. To address this and ensure a rigorous, privacy-preserving comparison, we reproduced their collection mechanism by strictly following the open-source logic of ByteDance MineContext but coupled it with the same on-device MLLM used in \tool. Specifically, this baseline employs a time-based strategy, capturing screen snapshots at fixed intervals to build a historical context. In our implementation, we set the sampling rate to 1 fps and fed the sequence directly to the local MLLM. 
    This setup isolates the impact of the sampling strategy, serving as a rigorous reference point to test the raw effectiveness of continuous monitoring without \compreduce.
    
    \item OCR-based Analysis: This baseline represents a strong, non-MLLM alternative. It first applies Optical Character Recognition (OCR) to frames captured by uniform sampling, and then uses a predefined set of risk-related keywords to flag risky events.
    
    \item \toolpf (Without Pixel-level Frame Reduction and Feature-level Compression): This baseline is the same as \tool, but without the pixel-level frame reduction (Step 2 of \compreduce) and feature-level compression (Step 3 of \compreduce). It is used to test the performance of action-guided adaptive sampling (Step 1 of \compreduce).
    
    \item \toolf (Without Feature-level Compression): This baseline is the same as \tool, but without the feature-level compression (Step 3 of \compreduce). It is used to test the performance of Step 1 and 2 of \compreduce.
\end{itemize}

\subsubsection{Evaluation Metrics}
To provide a comprehensive and rigorous assessment of \tool and the baselines, we define a set of metrics tailored to each research question.
For RQ1, we assess provenance fidelity from two aspects: the correctness of logging an event and the accuracy of labeling it.

\begin{itemize}%
    \item \textbf{Logging Correctness (Binary Classification):} To measure the fundamental ability to distinguish significant (risky) events from benign ones, we treat the problem as a binary classification task (Risk vs. No Risk). We report the Precision, Recall, and F1-Score for the positive (Risk) class. High Recall is particularly critical, as it indicates the system's ability to not miss potential threats. 
    
    \item \textbf{Labeling Accuracy (Multi-class Classification):} To evaluate the model's ability to correctly categorize the specific type of risk for a logged event, we report the macro-averaged F1-Score (Macro-F1) across all five categories. Macro-averaging ensures that performance on rare risk categories is weighted equally.
    \item \textbf{Description Quality (Logging Faithfulness):} To fully evaluate \tool's provenance fidelity, we introduce a human evaluation metric. For all 295 clips in our dataset, we measure the \textbf{Factual Correctness} of the MLLM-generated \texttt{abstractive\_summary}. This metric assesses whether the natural language log is an accurate, non-hallucinatory description of the visual evidence. We report the percentage of descriptions rated as factually correct versus those containing errors.
    \item \textbf{Evidence Volume (Average Number of Visual Tokens):} We calculate this metric to measure the effectiveness of \compreduce. It directly reflects the sampling and computational cost passed to the MLLM.
\end{itemize}

For RQ2, our evaluation of the query engine is divided into two parts, corresponding to the two primary query types. For event retrieval queries, we evaluate the accuracy in retrieving relevant events. We use standard metrics from the field of information retrieval. For a curated set of natural language questions, where each question is mapped to one or more ground-truth events in our dataset, we report the Mean Reciprocal Rank (MRR) and Recall@K. For detailed follow-up queries, we evaluate the accuracy in retrieving the specific detail.

\begin{itemize}%
    \item \textbf{Mean Reciprocal Rank (MRR):} MRR measures the average ranking quality for finding the first relevant event. It is particularly useful for evaluating how quickly a user can find at least one correct piece of evidence. 
    It is defined as:
    \begin{equation*}
        \text{MRR} = \frac{1}{|Q|} \sum_{i=1}^{|Q|} \frac{1}{\text{rank}_i}
    \end{equation*}
    where $Q$ is the set of questions, and $\text{rank}_i$ is the rank of the first relevant event for the $i$-th question.
    
    \item \textbf{Recall@K:} It measures the proportion of queries for which at least one relevant event is successfully found within the top K results. 
    We will report this for K=1 and K=3. 
    It is defined as:
    \begin{equation*}
        \text{Recall@K} = \frac{1}{|Q|} \sum_{i=1}^{|Q|} I(\exists j, \text{s.t.} (j \in \text{Res}_i(K) \wedge j \in \text{GT}_i))
    \end{equation*}
    where $I(\cdot)$ is an indicator function that is 1 if the condition is true and 0 otherwise, $\text{Res}_i(K)$ is the set of top- $K$ results for query $i$, and $\text{GT}_i$ is the set of all correct events for this query.
    
    \item \textbf{Accuracy:} We evaluate the accuracy of the retrieved event in answering the follow-up question.
\end{itemize}

\revise{For RQ3, we quantify the end-to-end on-device cost of the full \tool pipeline, including on-device MLLM inference, during a standardized usage session. We measure power consumption, battery temperature, per-event inference latency, peak memory footprint, and storage overhead across three modern smartphones.}

\subsection{RQ1: Provenance Fidelity}
\label{subsec:rq1}

\begin{table*}[t]
    \centering
    \caption{\revise{RQ1: Results for Provenance Fidelity evaluation. Precision, Recall, and F1-Score measure the binary quality of logging significant events (Risk vs. No Risk). Macro-F1 measures the accuracy of labeling those events across five categories. Avg. Tokens reflects the volume of visual evidence processed by the MLLM.}}
    \label{tab:rq1}
    \begin{tabular}{lccccc}
        \toprule
        \textbf{Method} & \textbf{Avg. Tokens} & \textbf{Precision} & \textbf{Recall} & \textbf{F1-Score} & \textbf{Macro-F1} \\
        \midrule
        OCR & - & 1.0000 & 0.4421 & 0.6131 & 0.4698 \\
        \tooluniform & 22850 & 0.9915 & 0.9957 & 0.9936 & 0.9798 \\
        \toolpf & 4792 & 0.9871 & 0.9871 & 0.9871 & 0.9561 \\
        \toolf & 3672 & 0.9872 & 0.9914 & 0.9893 & 0.9453 \\
        \tool & 2476 & 0.9830 & 0.9914 & 0.9872 & 0.9523 \\
        \bottomrule
    \end{tabular}
\end{table*}

To answer RQ1, we evaluate \tool's ability to faithfully create high-quality provenance records. We assess this from two primary aspects: the correctness of logging events and the accuracy of labeling them in the first place. The results are summarized in \autoref{tab:rq1}.

The \toolnaive (Naive Continuous Analysis) baseline is omitted from \autoref{tab:rq1}, as it failed to produce a valid result for the vast majority of our test cases. For an average 17-second clip, this method generates an average of 413 frames, which translates to over 270,000 visual tokens, far exceeding the context window capacity of current MLLMs and confirming that a brute-force approach is practically infeasible.

\noindent\textbf{Logging Correctness and Labeling Accuracy.} As shown in \autoref{tab:rq1}, our full \tool framework demonstrates high fidelity in labeling logged events, achieving a Macro-F1 Score of 95.23\%. This performance is highly competitive with the \tooluniform baseline (97.98\%). In contrast, the OCR-based baseline performs significantly worse, achieving a Macro-F1 Score of only 46.98\%. Though it gains a high precision of 100\%, the low recall of OCR-based baseline shows that it cannot identify a large amount of risky events. This is because its reliance on a predefined keyword set prevents it from understanding nuanced textual context, and furthermore, it is inherently incapable of interpreting either sequential actions or non-textual visual risks.

\begin{figure}[h]
    \centering
    \includegraphics[width=0.75\columnwidth]{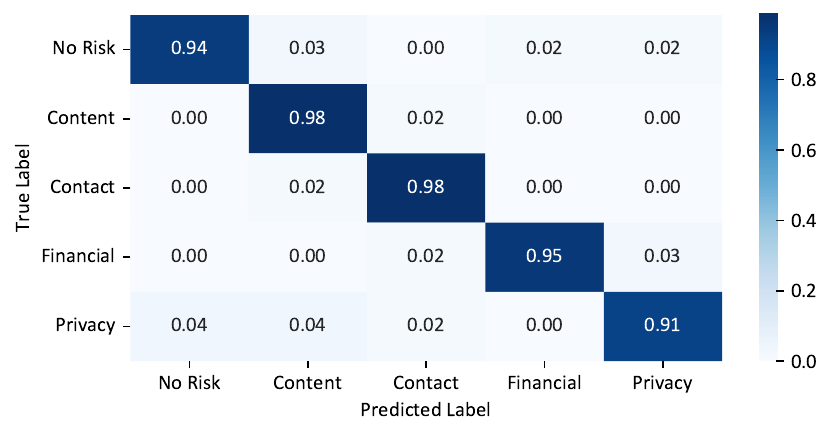}
    \caption{\revise{The confusion matrix of \tool's risk classification performance. The row represents the predicted risk categories, and the column represents the ground truth risk categories. The diagonal elements represent the number of correctly classified events, and the off-diagonal elements represent the number of misclassified events.}}
    \label{fig:confusion_matrix}
\end{figure}

To offer a more granular view, the confusion matrix in \autoref{fig:confusion_matrix} shows the labeling performance for each risk category. The fidelity is high (98\% Recall) for Content and Contact risks, and 95\% for Financial risks. It remains high but is slightly more challenged for Privacy risks (91\% Recall), as these threats are often defined by subtle contextual details, highlighting a promising direction for future work.
\revise{Furthermore, we split results by participant cohort: adult and underage participants achieved comparable Macro-F1 (94.65\% vs.\ 94.90\%), showing that the pipeline remains effective across both cohorts.}

\noindent\textbf{Evaluation of Description Quality (Logging Faithfulness).}
To fully validate the provenance fidelity (RQ1), we assessed not only classification accuracy (labeling) but also the factual correctness of the generated summaries. A faithful log requires an accurate natural language description, regardless of whether the event is classified as risky or benign. We conducted a human study on the \textbf{entire 295-clip dataset}, where 3 annotators rated the factual correctness of the MLLM-generated \texttt{abstractive\_summary} for each clip.

This exhaustive analysis showed high fidelity: \textbf{88.47\% (261/295)} of all descriptions were factually correct and non-hallucinatory. The remaining 34 clips contained descriptive errors, which we categorized as follows:
\begin{itemize}%
    \item \textbf{Logging-Only Errors (6.44\%):} The description correctly identified the risk, but made a minor factual error (\eg, ``the user is listening to music'' when they were watching a post containing a screenshot of a music album on social media).
    \item \textbf{Classification-Driven Errors (5.08\%):} The description was flawed \textit{because} the MLLM fundamentally misclassified the event (\ie, these errors correspond to the misclassifications captured by our Macro-F1 score).
\end{itemize}

This result confirms that \tool not only labels risks accurately (Macro-F1) but also logs the events with high fidelity, further validating the provenance fidelity of our system.

\noindent\textbf{Efficiency of \compreduce.} The critical insight, however, lies in the efficiency dimension. While achieving comparable effectiveness to periodic snapshot, \tool does so by processing an average of only 2,476 tokens per clip. This represents a data reduction of over 89.2\% compared to the 22,850 tokens required by the \tooluniform baseline for the same video clips. This result powerfully demonstrates that our hierarchical, event-driven distillation pipeline achieves a vastly superior fidelity-to-cost tradeoff.

\noindent\textbf{Ablation Study on Distillation Steps.} The effectiveness of each step in our distillation process is validated by our ablation studies. Starting from \tooluniform, the action-guided adaptive sampling step alone (\toolpf) achieves the largest token reduction (79.0\%). Subsequent steps (fast deduplication and semantic compression) further reduce the token count while the Macro-F1 score remains remarkably stable (hovering around 95\%), proving that each stage effectively removes redundant data without sacrificing essential semantic signals.

\revise{
\noindent\textbf{Sensitivity Analysis on Sampling Design.} Our selected configuration ($F_{high}=2$ fps, $t_{window}=2$s) sits at the optimal fidelity-to-cost trade-off, achieving a Macro-F1 of 95.23\% with an average token cost of 2,476. We verified this by varying $F_{high}$ and $t_{window}$ separately and reporting both Macro-F1 and average token count for each setting (\autoref{fig:sensitivity}).

With $t_{window}$ fixed at 2s, dropping $F_{high}$ to 1 fps causes a Macro-F1 drop of 8.42\% (to 86.81\%), as the system occasionally misses brief transition states or animations. Raising $F_{high}$ to 4 fps yields only a 0.33\% gain (to 95.56\%) but roughly doubles the token overhead to over 4,100. Similarly, with $F_{high}$ fixed at 2 fps, a 1s window truncates evidence for slower-loading pages (Macro-F1 drops by 6.10\%, to 89.13\%), while 3s or 4s windows inflate the token count (to 3,250 and 4,080, respectively) with no proportional accuracy gain (Macro-F1 plateaus around 95.6\%).

\begin{figure}[h]
    \centering
    \includegraphics[width=1\columnwidth]{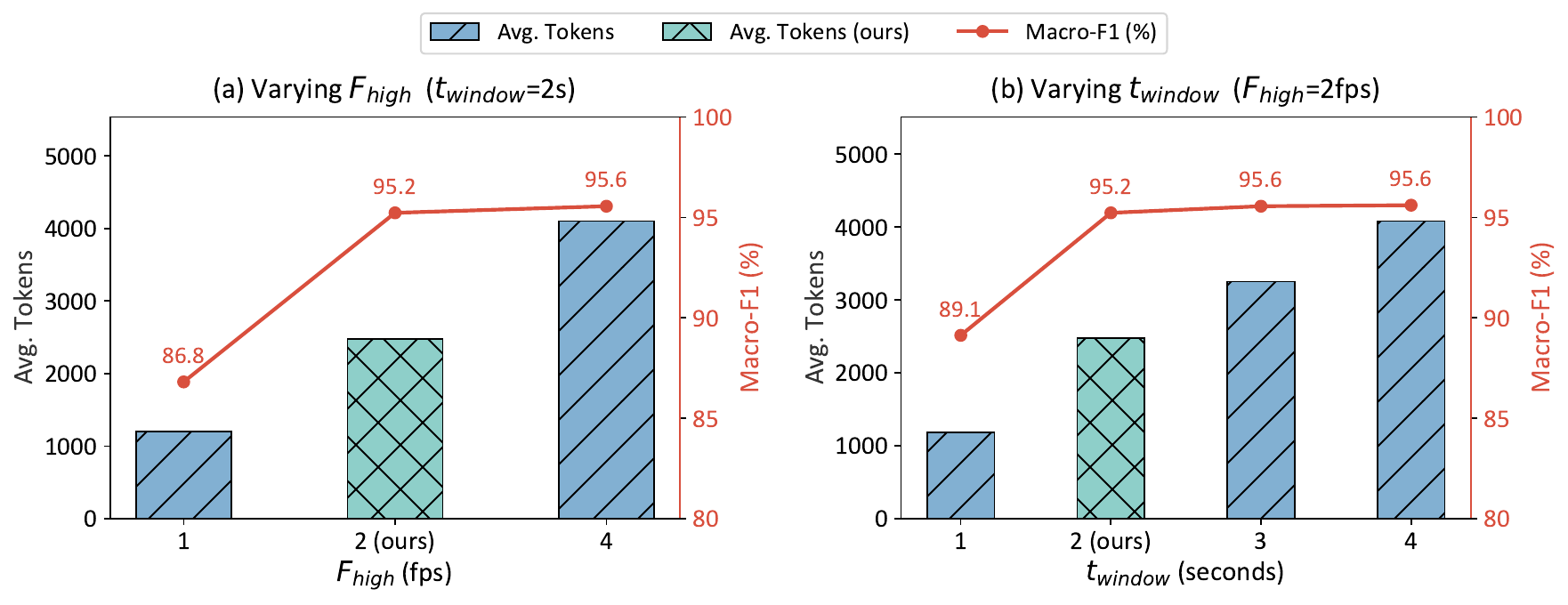}
    \caption{\revise{Sensitivity analysis of the two core sampling parameters. (a)~Varying the sampling rate $F_{high}$ with $t_{window}$ fixed at 2s. (b)~Varying the capture window $t_{window}$ with $F_{high}$ fixed at 2\,fps. Bars show the average visual tokens per clip (left axis); the line shows Macro-F1 (right axis). The green bar marks our selected configuration.}}
    \label{fig:sensitivity}
\end{figure}
}

\revise{\noindent\textbf{Per-Action Frame Retention Analysis.} Analyzing frame retention at the interaction level further explains why \tool remains efficient under continuous use. On average, a 2-second window at 2 fps generates exactly 4 candidate frames per action. Step 2 (Fast Deduplication via Pixel Similarity) removes redundant frames caused by UI rendering delays or rapid tapping, keeping an average of 1.6 visually distinct frames per action. Step 3 (Semantic Intra-frame Compression) then isolates the fine-grained changes, passing only 18.5\% of the original visual token volume to the MLLM. Even under rapid, continuous interaction bursts, the downstream deduplication stages keep the per-event MLLM inference cost bounded.
}

\begin{tcolorbox}[size=small, title=Answer to RQ1,breakable]
    \tool demonstrates high provenance fidelity. It accurately logs significant events with a 98.72\% F1-Score and correctly labels their risk type with a 95.23\% Macro F1-Score\revise{, with comparable performance across adult and underage cohorts}. It achieves this fidelity while reducing the volume of visual evidence requiring analysis by over 89.2\% compared to the prevailing industrial periodic snapshot strategy (MineContext), making high-fidelity auditing viable on mobile batteries. The generated logs are also faithful: 88.47\% of the descriptions were factually correct and non-hallucinatory. \revise{A sensitivity analysis shows our sampling configuration ($F_{high}=2$ fps, $t_{window}=2$s) sits at the optimal fidelity-to-cost trade-off. Per-action analysis shows downstream filtering retains only 1.6 distinct frames and 18.5\% of tokens per action.}
\end{tcolorbox}

\subsection{RQ2: Forensic Capability}

\noindent \textbf{Methodology.} To evaluate the forensic capability of \tool's query engine, we created a test set of 68 natural language questions. To ensure these questions are both representative of real-world concerns and methodologically sound, we first form a pool of potential investigation topics based on the parental concerns identified in our Stage 1 interviews (\S~\ref{subsubsec:data_construction}). Three researchers, who were not involved in the system's development, then independently drafted, cross-validated, and refined this pool into a final, curated test set of 51 retrieval and 17 interrogation questions, each of which was then manually mapped to one or more ground-truth events in our dataset. \revise{The queries were designed to reflect realistic parental inquiry patterns, not simply mirror our predefined risk categories. The set includes category-aligned queries (\eg, ``Did my child make any purchases?'') as well as cross-category and open-ended queries that require semantic understanding (\eg, ``Did my child interact with any strangers?'', ``Was there anything inappropriate in what my child watched today?''). This mix tests semantic retrieval capability rather than simple category matching.}

\noindent \textbf{Results.} The evaluation of our query engine was divided into two parts to separately assess its two primary forensic capabilities, with results summarized in \autoref{tab:rq2}.

\begin{table}[h]
    \centering
    \caption{\revise{RQ2: Results for Forensic Capability evaluation. Recall@K (K=1,3) and MRR evaluate the performance of Stage 1 (Event Retrieval) on a set of 51 natural language questions. Accuracy (Acc.) evaluates the performance of Stage 2 (Event Interrogation) on a separate set of 17 follow-up questions.}}
    \label{tab:rq2}
    \begin{adjustbox}{width=0.5\columnwidth}
    \begin{tabular}{lcccc}
        \toprule
        & Recall@1 & Recall@3 & MRR & Acc. \\
        \midrule
        \tool & 0.9020 & 0.9412 & 0.9234 & 0.8824 \\
        \bottomrule
    \end{tabular}
    \end{adjustbox}
\end{table}

For event retrieval performance, the first set of 51 questions was designed to test the engine's ability to retrieve the correct event(s) from the entire dataset. \tool achieved a MRR of 0.9234, indicating that for the vast majority of queries, the correct relevant event was ranked at or very near the top of the results. Furthermore, \tool achieved a Recall@1 of 90.20\% and a Recall@3 of 94.12\%. This demonstrates that not only is the correct event almost always retrieved within the top three results, but it is the single best result in over 90\%+ of cases. These strong results confirm the effectiveness of our hybrid storage and two-stage forensic query engine for practical, in-depth forensic investigation.

For event interrogation performance, the second set of 17 questions was designed to test the engine's ability to answer detail-oriented, follow-up questions about a specific, pre-identified event. \tool achieved an accuracy of 88.24\%, successfully extracting specific details (\eg, monetary amounts, usernames) from the stored visual evidence.

\noindent\textbf{Failure Cases.} To better understand the system's limitations, we delved into the qualitative failure cases from our query evaluation. A representative failure occurred in a content risk scenario. While browsing a social media feed, the child encountered a post containing a GIF of frying pork accompanied by inflammatory text. In this specific context, the combination of the visual and text was intended as a derogatory religious metaphor to mock Muslims.  In this case, \tool's MLLM correctly logged the event as a Content Risk during the initial triage. However, its generated \texttt{abstractive\_summary} was superficial (``The user viewed a GIF of food being cooked'') and failed to capture the malicious religious context, causing a subsequent semantic query for religious related risks to fail. This failure case is a direct, qualitative example of the ``Logging-Only Errors'' (6.44\%) that we quantified in \S~\ref{subsec:rq1}. It demonstrates precisely why logging fidelity is critical: a semantically poor description, even with a correct label, directly undermines the guardian's ability to perform post-hoc forensic queries.

This case highlights two important findings. First, it underscores a known limitation of current MLLMs in comprehending complex, culturally-specific metaphors, further demonstrating the inherent fallibility of a fully autonomous approach. Simultaneously, it showcases the effectiveness of our architecture: although the semantic query failed, the underlying evidence was faithfully preserved and remained accessible to the guardian via broader queries, validating our human-in-the-loop design. Second, this reveals a clear direction for future work in enhancing the system's capabilities, such as improving the MLLM's domain-specific reasoning about implicit and metaphorical risks.

\begin{tcolorbox}[size=small, title=Answer to RQ2,breakable]
    \tool's two-stage query engine demonstrates strong forensic capabilities. It excels at retrieving the correct event record in response to broad natural language questions (over 90\% Recall@1) and is highly accurate in extracting specific details from visual evidence for follow-up questions (88.24\% accuracy). This confirms the effectiveness of our design for enabling practical, human-led investigation.
\end{tcolorbox}
\revise{
\subsection{RQ3: On-Device Performance Overhead}
\label{subsec:rq3_overhead}

\revise{We deploy \tool, \tooluniform, and \toolnaive on the three smartphones described in \S~\ref{subsubsec:setup} and measure the end-to-end on-device overhead with the 4-bit quantized model. Following the same evaluation setup as RQ1--RQ2, we ran each method through a continuous 1-hour session covering all interaction scenarios (browsing social media, watching short videos, chatting, and performing in-app transactions).} Power consumption was collected using Battery Historian~\cite{battery-historian}, which aggregates readings across all hardware components from \texttt{/sys/class/power\_supply}~\cite{xu2022mandheling}. Battery temperature was read from \texttt{/sys/class/thermal/thermal\_zone*/temp} at one-second intervals and averaged over the session. \revise{Each method was tested five times on each device; we report cross-device averages as deltas over normal phone use. We also measured the storage footprint of the provenance records produced by \tool. The results are shown in \autoref{fig:rq3}.}

\begin{figure}[h]
    \centering
    \includegraphics[width=1\columnwidth]{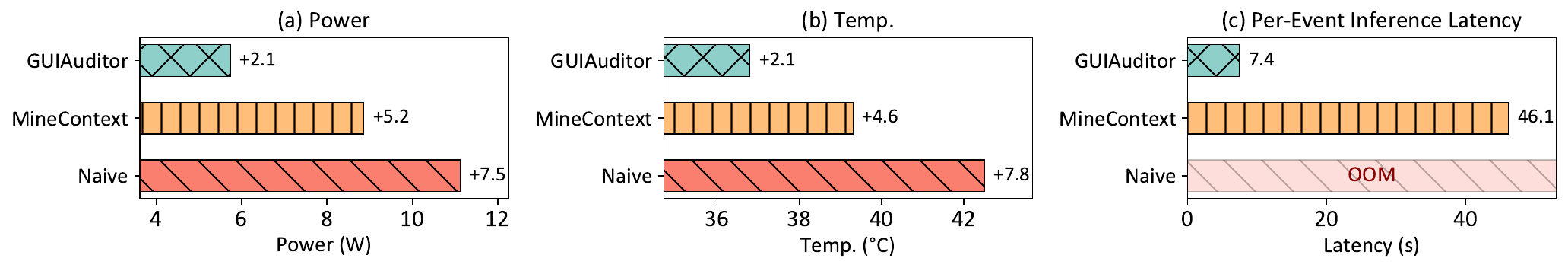}
    \caption{\revise{End-to-end on-device overhead over a 1-hour session, averaged across three modern smartphones. All values are deltas over normal phone use. (a)~Power consumption. (b)~Battery temperature. (c)~Per-event inference latency (\toolnaive triggers OOM).}}
    \label{fig:rq3}
\end{figure}

\revise{\noindent\textbf{Power Consumption.} Under our one-hour mixed-use workload, \tool adds 2.1W over normal phone use, corresponding to roughly 10\% additional battery drain per hour on the tested devices. Given that children's daily screen time typically ranges from 2 to 4 hours~\cite{anderson2023teens}, this measured overhead suggests that post-hoc auditing is manageable on modern smartphones, though longer field deployments may expose additional usage patterns. In contrast, \tooluniform draws 5.2W (2.5$\times$ of \tool), draining approximately 25\% per hour, and \toolnaive draws 7.5W (3.6$\times$), draining approximately 36\% per hour, making them less suitable for sustained use.}

\revise{\noindent\textbf{Battery Temperature.} \tool raises the device temperature by only 2.1$^\circ\text{C}$, well below the skin-temperature throttling threshold of modern smartphones~\cite{pu2023investigating}. \tooluniform and \toolnaive raise temperatures by 4.6$^\circ\text{C}$ (2.2$\times$) and 7.8$^\circ\text{C}$ (3.7$\times$), respectively. Temperature increases of this magnitude trigger thermal throttling on mobile SoCs, forcing CPU frequency reductions that degrade both inference throughput and user experience~\cite{zhou2022play}.}

\revise{\noindent\textbf{Inference Latency and Memory.} \tool completes each event-triggered MLLM invocation in 7.4s on average and uses 3.1GB peak memory. Since \tool is post-hoc, inference runs after the detected interaction and does not block the child's device use. Under extremely dense event bursts, however, even \tool may build a temporary inference backlog. This would delay when a guardian can inspect the corresponding records, not the child's ongoing interaction. \tooluniform requires 46.1s per event (6.2$\times$) and 6.7GB peak memory (2.2$\times$), mainly because uniform sampling produces many more visual tokens. During active use, this latency can cause queued events to accumulate. \toolnaive cannot complete inference: its per-invocation token volume exceeds available device RAM, so the OS repeatedly kills the runtime.}

\revise{\noindent\textbf{Storage Overhead.} \tool's storage footprint is negligible relative to modern device capacity. In our 1-hour session, \tool logged an average of 113 events across devices. Each provenance record consists of a text summary (1.4KB), an embedding vector (3KB), and a small set of distilled keyframes (47.3KB), totaling 51.7KB per record. The 113 records per hour occupy 5.9MB. Extrapolating to 2--4 hours of daily screen time~\cite{anderson2023teens}, the daily storage overhead is 11.8--23.6MB even before discarding No Risk records, which is negligible on devices with 256GB or more of storage. As records accumulate over weeks or months, retrieval efficiency can be maintained with standard approximate nearest-neighbor indexing (\eg, HNSW~\cite{malkov2018efficient}) and time-based partitioning. In deployment, guardians can configure a retention window (\eg, 30 or 90 days) to automatically purge older records, bounding long-term storage growth.}

\begin{tcolorbox}[size=small, title=Answer to RQ3,breakable]
    \revise{\tool runs with manageable end-to-end overhead on three modern smartphones under our one-hour mixed-use workload. It adds roughly 10\% battery drain per hour and stores only 5.9MB/hr of provenance records. Its resource use is also substantially lower than both \toolnaive, which exceeds device RAM and cannot complete inference, and \tooluniform, which incurs 2--6$\times$ the overhead across the measured metrics. These results show that event-driven distillation makes post-hoc auditing feasible on the tested mobile hardware.}
\end{tcolorbox}
}

\section{Discussions}
\revise{This section discusses the implications of \tool's design for child-safety forensics. We explain the need for a human-in-the-loop paradigm, examine risks of coercive control and privacy/data-security considerations, and close with limitations and future work.}

\revise{\subsection{Necessity of the Human-in-the-Loop Paradigm}
\label{subsec:hitl_necessity}

The post-hoc, human-in-the-loop design of \tool is a deliberate choice, not a workaround for imperfect models. It addresses two practical needs that fully automated systems cannot satisfy: guardians need to understand what their children experience online, and children need to use their phones without disruption.

\noindent\textbf{Guardian Perspective.}
When a child encounters something harmful online, a guardian's first concern is not just stopping it but understanding what happened. Did the child stumble onto the content by accident, or did a recommendation algorithm push it? Was the child confused, frightened, or indifferent? These questions matter because they determine how the parent should respond, whether to have a conversation, adjust device settings, or seek further help~\cite{wang2021protection}. Existing detection-based systems cannot answer any of them. They block content or terminate sessions, but the guardian only learns that \textit{something} was blocked, with no context about how the situation arose or how the child reacted. \tool's provenance records fill this gap by preserving the captured interaction context so that guardians can review events and make informed decisions.

\noindent\textbf{Child Perspective.}
Children today rely on their phones for everyday tasks: buying meals, hailing rides, looking up schoolwork, and communicating with friends~\cite{wang2021protection,ghosh2018safety}. Maintaining this normal functionality is, in most situations, more important than blocking a potentially risky action, because the vast majority of a child's phone interactions are benign and necessary. Real-time blocking inevitably produces false positives, and a single wrongly blocked action can prevent a child from completing these routine activities. Frequent interventions also make children feel surveilled, which erodes trust and motivates workarounds like borrowing a friend's device or creating unmonitored accounts~\cite{akter2022parental}. A post-hoc system avoids blocking the child's interaction while still allowing the guardian to review flagged events afterward.

\noindent\textbf{Relationship to Model Capability Improvements.}
We note that \tool's capability grows with the underlying MLLM: as models become more capable, the risk classifications and semantic summaries fed to guardians become more accurate and more detailed. At the same time, the human-in-the-loop design is orthogonal to model capability. As argued above, guardians need contextual understanding of incidents to respond appropriately, and children need their device use to remain uninterrupted. Neither need is a consequence of current model limitations, and neither disappears with a stronger model.

\tool is intended to \textit{complement}, not replace, real-time defenses such as app blockers, content filters, and transaction-level safeguards. It lets guardians discover threats that those tools miss and verify their alerts using the captured interaction context. Immediate harms that can occur within a single session, such as unauthorized payments, still require preventive safeguards at the OS, app, or payment layer. Because \tool's analysis runs asynchronously, per-event inference latency (\S~\ref{subsec:rq3_overhead}) does not block the child's interaction, although dense event bursts may delay when records become available to the guardian.
}

\subsection{Risk of Coercive Control}
Like all parental control technologies, \tool operates within complex interpersonal dynamics and inevitably faces the dual-use risks common to this domain. While our design is predicated on informed consent, we acknowledge that there may still be risks in scenarios involving coercive control, particularly within intimate partner relationships or imbalanced parent-child dynamics. In such contexts, a tool designed for protection can be weaponized as an instrument of oppression. The demand for installation becomes a loyalty test, and refusal can be manipulated as evidence of wrongdoing, effectively nullifying the notion of voluntary consent.

While solving this deep societal problem extends beyond the scope of a single technical paper, we argue that responsible system design must actively incorporate mitigations to raise the bar against misuse. As critical directions for future work, we propose technical barriers to strictly enforce the intended parent/minor relationship at setup and deter misuse for partner surveillance. For example, we could leverage a facial camera to detect whether the monitored phone belongs to a child and the parent-child relationship is real~\cite{akter2023takes,akter2022parental}. We can also rely on methods that incorporate anti-coercion features to further strengthen the identification~\cite{cai2024famos}. Furthermore, behavioral analysis to detect obsessive monitoring~\cite{bellini2023digital}, combined with a tamper-proof, immutable audit log for the monitored user~\cite{havron2019clinical,ahmad2022hardlog,jiang2025dpuaudit}, would be essential for accountability. Finally, these technical measures must be complemented by explicit policy enforcement: the tool's positioning as solely a child safety aid must be reinforced through user agreements and prominent UI warnings that expressly forbid misuse like partner surveillance and articulate potential legal consequences.

\subsection{Privacy and Data Security Considerations}
\label{subsec:privacy}
The deployment of a monitoring system like \tool, especially for children, requires paramount attention to privacy and data security~\cite{ghosh2018safety,koh2025harnessing}. Our design incorporates several measures to address these concerns.

First, the deployment of \tool is predicated on the explicit, informed consent of the guardian, positioning it as a tool for parental oversight, not covert surveillance. Importantly, children themselves should also be involved in this process~\cite{fu2022social}. While younger children may not be able to fully articulate informed consent, involving them in age-appropriate explanations strengthens trust and avoids creating a purely paternalistic model. For older minors, explicit acknowledgement of the system's use aligns with international guidelines such as the GDPR's emphasis on fairness and transparency~\cite{voigt2017eu}.

Second, as a core design choice, \tool is architected so that all analysis is performed exclusively on-device, with no screen content ever transmitted to an external server. \revise{Our end-to-end evaluation (\S~\ref{subsec:rq3_overhead}) shows that the full pipeline, from event detection and evidence distillation to 4-bit quantized Qwen2.5-VL-3B inference, runs natively on three modern smartphones under our measured workload with manageable overhead. As a result, screen content does not need to leave the device.} Our system only persists the minimal set of core visual evidence for significant events, which directly reflects the GDPR requirement that collected data must be adequate, relevant, and limited to what is necessary~\cite{voigt2017eu}.

\revise{Third, even a well-intentioned monitoring tool can infringe on a child's developing sense of privacy and autonomy. \tool's design limits this risk in three ways: (1) it does not record continuously but only captures evidence around discrete interaction events; (2) it discards all data for events classified as ``No Risk'', so benign activities leave no trace; and (3) the guardian interacts with structured summaries and targeted queries rather than browsing raw screen recordings. Transparent deployment, where children are aware of and involved in the system's use (as discussed above), is also necessary for balancing safety with the child's privacy.}

Fourth, the on-device provenance records can be further hardened with standard mechanisms, such as file-system-level encryption~\cite{delaune2024formal} or running analysis within a TEE~\cite{moon2025asgard,zhang2024no,wanggame}, though such hardening is orthogonal to our core contribution.

\subsection{Limitations and Future Work}
\label{subsec:limitations_future}

\noindent \textbf{Reliance on MLLM Capabilities.} The quality of the semantic summaries in \tool's provenance records depends on the reasoning capabilities of the underlying MLLM. While these models are effective for general analysis, they can miss complex, implicit, or culturally specific cues. This was evident in our failure case analysis: the MLLM correctly identified an event as a Content Risk but produced a superficial summary that overlooked a malicious religious metaphor. However, this limitation only affects the initial semantic abstraction. The underlying evidence is still preserved, so a guardian can retrieve the event via broader queries (\eg, by time or risk type) and use the Stage 2 interrogation engine to ask detailed questions about the visual content.

\revise{\noindent\textbf{Recruitment Difficulty and Dataset Scale.}
Recruiting minors for research involving sensitive screen-recorded interactions is difficult in practice. Guardians are understandably cautious about consenting to such studies, and younger children are often uninterested or uncooperative. We contacted over 1,000 potential participants and their families; ultimately only 32 (< 3\%) agreed to participate. The rationale for our participant composition and age coverage is discussed in detail in \S~\ref{subsubsec:data_construction}. While a larger and younger cohort would be desirable, the current dataset of 295 clips across 32 participants already spans the legally significant age bands (under 14 per China's PIPL~\cite{pipl2021}, under 16 per the EU GDPR~\cite{voigt2017eu}) and shows comparable results across adult and underage cohorts (\S~\ref{subsec:rq1}). Given the difficulty of recruiting minors for this type of study, this dataset provides a practical level of age coverage for the current evaluation.}

\revise{\noindent\textbf{Risk Taxonomy Coverage.}
Our taxonomy, derived from the 4Cs framework (\S~\ref{subsec:taxonomy}), covers the established risk types documented in governmental and regulatory reports. The taxonomy does not claim to be exhaustive: new risk types will emerge as platforms evolve, and real-world risks may not fall neatly into a single category (\eg, grooming can simultaneously involve contact risk and privacy violation). Our scenarios and RQ2 queries were partly organized around this taxonomy, so they may still reflect the structure of our predefined categories. We reduced this bias by deriving scenarios from parent interviews and public reports, asking annotators to label clips without seeing task scripts, and constructing queries from parental concerns rather than category names alone. Still, scripted tasks and researcher-curated queries cannot fully capture longitudinal, in-the-wild use or the full diversity of questions guardians may ask in deployment. However, adapting \tool to a new category is lightweight. Because classification relies on chain-of-thought prompting rather than a trained classifier, adding a new risk type only requires updating the prompt template, with no model retraining or data collection. Even when the MLLM's classification is imprecise, the underlying evidence is preserved for the guardian to judge via Stage 2 interrogation.}

\section{Related Work}

\noindent \textbf{Child Online Safety.} A significant body of research uses qualitative methods to understand the landscape of online risks children face across various emerging domains, from vulnerabilities in parental control apps themselves to safety concerns in Virtual Reality and Generative AI~\cite{ali2020betrayed,cao2024understanding,yu2025exploring}. While this body of work is invaluable for establishing the problem's severity from a sociological or HCI perspective, our contribution is orthogonal as we focus on designing and building a technical system to enable the post-hoc forensic analysis of these risks.

\noindent \textbf{GUI Automation and Testing Agents.} Recent work uses LLMs as agents to automate GUI tasks or find bugs~\cite{liu2024make,wen2024autodroid,wen2024autodroid2,cuadra2024digital}. While we also interpret GUIs, our objective is fundamentally different: those works prioritize task completion by an agent, whereas \tool audits a human's interactions for safety. This leads to a different technical challenge: prioritizing the creation of a faithful, verifiable record for human-led investigation.

\section{Conclusion}
\label{sec:conclusion}
In this paper, we address the insufficiency of fully automated detection for protecting children from complex in-app risks by introducing a human-in-the-loop, post-hoc forensic paradigm. We present \tool, the first system to materialize this vision through the core concept of GUI Provenance: a structured, queryable record of a user's interaction sequence. By translating the raw GUI stream into this new form of verifiable evidence, \tool gives guardians contextual evidence that automated systems often lack. A comprehensive evaluation on a new benchmark shows that our design, featuring an evidence distillation pipeline and a two-stage query engine, supports accurate forensic investigation with moderate overhead on the tested mobile devices. \revise{An end-to-end evaluation with a 4-bit quantized MLLM on three modern smartphones shows that the full pipeline, from evidence distillation to on-device inference, runs on current hardware under our measured workload.} This work lays the foundation for a new generation of security tools that assist and empower, rather than replace, the human guardian.

\begin{titlecasesections}
\section*{Acknowledgments}
We sincerely thank the editors and reviewers for their valuable comments and suggestions, which greatly enhanced the quality of our paper. Correspondence should be addressed to Yifeng Cai, Ding Li, and Yao Guo.
This work was partly supported by the Beijing Natural Science Foundation (L243010), the National Natural Science Foundation of China (U25A6024), and the Fundamental and Interdisciplinary Disciplines Breakthrough Plan of the Ministry of Education of China (JYB2025XDXM108).
\end{titlecasesections}

\bibliographystyle{ACM-Reference-Format}
\begin{titlecasesections}
\bibliography{ref}
\end{titlecasesections}
\section{Appendix}

\subsection{Declaration of Generative AI Usage}
During the preparation of this work, the authors used ChatGPT 4.5 and 5.1 to improve the quality of writing, including style, phrasing, and grammar. The prompts used were strictly limited to editorial requests (\eg, ``proofread this paragraph'', ``fix grammatical errors'') and did not involve the generation of new scientific concepts, text, or other content. All AI-generated outputs were rigorously inspected and refined by the authors to ensure accuracy and originality. The authors maintain full responsibility for the entire content of this manuscript.

\end{document}